\documentclass{IEEEoj}
\ifCLASSINFOpdf
   \usepackage[pdftex]{graphicx}
   \DeclareGraphicsExtensions{.pdf,.jpeg,.png}
\else
   \usepackage[dvips]{graphicx}
   \DeclareGraphicsExtensions{.eps,.pdf}
\fi
\usepackage{mathtools}
\usepackage{amssymb}
\usepackage{amsthm}
\usepackage{cancel}
\usepackage{breqn}
\usepackage{multirow}
\usepackage{rotating}
\usepackage{verbatim}
\usepackage{textcomp,gensymb}
\usepackage{ushort}
\usepackage{caption}
\usepackage{algorithm}
\usepackage{siunitx}
\usepackage[noend]{algpseudocode}

\usepackage[backend=biber,style=ieee,doi=false,isbn=false]{biblatex}
\ifCLASSOPTIONcompsoc
    \usepackage[caption=false, font=normalsize, labelfont=sf, textfont=sf]{subfig}
\else
\usepackage[caption=false, font=footnotesize]{subfig}
\fi

\newcommand{\Expect}{{\rm I\kern-.5em E}}

\DeclareMathAlphabet{\mathitsf}{\encodingdefault}{\sfdefault}{m}{sl}
\DeclareMathAlphabet{\mathitbfsf}{\encodingdefault}{\sfdefault}{bx}{sl}
\def\vec#1{\ushortw{\boldsymbol{#1}}}
\def\matrix#1{\ushortdw{\boldsymbol{#1}}}
\DeclareMathOperator{\tr}{tr}
\newcommand{\rvline}{\hspace*{-\arraycolsep}\vline\hspace*{-\arraycolsep}}
\newcommand{\diff}{\mathop{}\!\mathrm{d}}
\usepackage{cool}
\newtheorem{theorem}{Theorem}[section]
\newtheorem{corollary}{Corollary}[theorem]
\newtheorem{lemma}[theorem]{Lemma}
\newtheorem{definition}{Definition}[theorem]
\renewcommand{\Re}{\mathrm{Re}}
\renewcommand{\Im}{\mathrm{Im}}

\def\BibTeX{{\rm B\kern-.05em{\sc i\kern-.025em b}\kern-.08em
		T\kern-.1667em\lower.7ex\hbox{E}\kern-.125emX}}
\AtBeginDocument{\definecolor{ojcolor}{cmyk}{0.93,0.59,0.15,0.02}}
\def\OJlogo{\vspace{-14pt}\includegraphics[height=28pt]{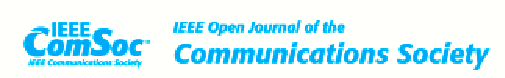}}
\begin{document}
	\receiveddate{XX Month, XXXX}
	\reviseddate{XX Month, XXXX}
	\accepteddate{XX Month, XXXX}
	\publisheddate{XX Month, XXXX}
	\currentdate{XX Month, XXXX}
	\doiinfo{OJCOMS.2023.1234567}

\title{Polarization-Based Security: Safeguarding Wireless Communications at the Physical Layer}

\author{POL HENAREJOS\authorrefmark{1}, SENIOR, IEEE, ANA I. PÉREZ-NEIRA\authorrefmark{1},
	FELLOW, IEEE}
\affil{Centre Tecnològic de Telecomunicacions de Catalunya (CTTC), Castelldefels, 08860, Barcelona, Spain}
\corresp{CORRESPONDING AUTHOR: Pol Henarejos (e-mail: pol.henarejos@cttc.es).}
\authornote{This work was supported by the Spanish Ministry of Economy and Competitiveness (Ministerio de Economia y Competitividad) under project IRENE (PID2020-115323RB-C31).}
\markboth{Preparation of Papers for IEEE OPEN JOURNALS}{Pol Henarejos \textit{et al.}}

\maketitle

\begin{abstract}
	Physical layer security is a field of study that continues to gain importance over time. It encompasses a range of algorithms applicable to various aspects of communication systems. While research in the physical layer has predominantly focused on secrecy capacity, which involves logical and digital manipulations to achieve secure communication, there is limited exploration of directly manipulating electromagnetic fields to enhance security against eavesdroppers. In this paper, we propose a novel system that utilizes the Mueller calculation to establish a theoretical framework for manipulating electromagnetic fields in the context of physical layer security. We develop fundamental expressions and introduce new metrics to analyze the system's performance analytically. Additionally, we present three techniques that leverage polarization to enhance physical layer security.
\end{abstract}



\section{INTRODUCTION}
\IEEEPARstart{P}{hysical} layer security, a discipline with roots in the ancient practice of cryptography, has played a crucial role in human communication throughout history. Early civilizations like the Spartans and Romans employed rudimentary encryption systems, laying the groundwork for what is now known as physical layer security. Significant advancements during World War II, such as the Enigma machine, propelled the field forward. However, the modern era of physical layer security began with Claude Shannon's influential paper \cite{Shannon1945,Shannon1949}. Since then, computer encryption has been the predominant foundation of physical layer security, encompassing symmetric key algorithms like Data Encryption Standard (DES) \cite{Biryukov2011} and Advanced Encryption Standard (AES) \cite{AES2001}, as well as asymmetric key systems like Rivest-Shamir-Adleman (RSA) \cite{RSA78} and Elliptic Curve Cryptography. However, the looming threat of quantum computing poses new challenges to their unbreakability.

Nevertheless, the existing literature on the physical layer security of the Physical Layer (PHY) remains limited. Shannon introduced the concept of \emph{secrecy capacity} \cite{Shannon1949}, which defines the region of capacity where an eavesdropper cannot extract information. Many works focus on increasing the capacity of the legitimate receiver against eavesdroppers to enhance secrecy \cite{4626059,6930805}, while others employ pre-distortion techniques to make the transmitted signal comprehensible only to eavesdroppers unaware of the employed filter \cite{8617993}.

Proposed systems delve deeper into the PHY, employing digital manipulations at the symbol level and utilizing both symmetric and asymmetric techniques \cite{8681520}. For example, the introduction of chaotic modulations enables the implementation of physical layer security systems \cite{8723465,8584431}.

Despite these digital manipulations at the PHY, electromagnetic waves are still transmitted using common and well-known modulations, making them susceptible to recording and subsequent offline exploitation by forensic systems for decryption. A simple example is the demodulation of a QPSK signal, enabling an eavesdropper to recover encrypted information through brute force attacks.

While quantum communications, such as Quantum Key Distribution (QKD), have emerged as a notable approach, exploiting the quantum properties of electromagnetic fields to transmit sensitive information securely, they are limited to specific applications and require high-precision instruments. In contrast, this paper focuses on a general approach for non-quantum communication systems, enhancing the overall robustness of end-to-end transmission.

This paper aims to explore the concept of physical layer security in its purest form: manipulating electromagnetic fields according to secret patterns based on a secret key to obfuscate the received electromagnetic wave from eavesdroppers. Unlike secrecy capacity systems, the goal is not to isolate the eavesdropper's capacity. In the proposed scheme, capacity is irrelevant in terms of physical layer security, as it only determines the maximum achievable rate.

Currently, there is limited research on the physical layer security of electromagnetic waves. For instance, Ghost Polarization Communication (GPC) \cite{GPC2020} utilizes unpolarized light beams and a secret reference to transmit encrypted information. However, this system is incompatible with polarized communication systems, such as common radio communications, and its bit rate falls short of meeting the demands of modern society.

In this paper, we leverage the polarization state to obfuscate the electromagnetic wave by employing an extensive Mueller calculus. Through the application of Mueller calculus, we can manipulate the physical polarization state in a manner that prevents eavesdroppers from extracting any information from the received electromagnetic field. While the Mueller calculus is not a novel theory \cite{Xing1992,Gil2000}, it has not been previously utilized for manipulating electromagnetic fields in the context of physical layer security.

To contextualize our work, we remark the following novel contribution:
\begin{enumerate}
	\item We introduce the Mueller calculus for cryptography in the PHY layer.
	\item We present a set of Theorems and Lemmas that describe analytically the conditions to achieve a secure cryptography. 
	\item We describe the encryption and decryption processes of the polarization.
	\item We study analytically the statistics of the received signal.
	\item We present two novel metrics to measure the strength of any cryptographic system in the PHY layer.
	\item We propose three different cryptographic systems.
	\item We analyze the impact of imperfect polarization in these cryptographic systems.
\end{enumerate}

\emph{Notation}: we denote scalars as $a$, vectors as $\vec{a}$, matrices as $\matrix{A}$, trace of a matrix as $\tr\left(\matrix{A}\right)$, determinant of a matrix as $\det\left(\matrix{A}\right)$, complex conjugate as $a^*$, the inverse of a matrix as $\matrix{A}^{-1}$, the conjugate of a matrix as $\matrix{A}^*$, the transpose of a matrix as $\matrix{A}^T$, the hermitian of a matrix as $\matrix{A}^H$, the expectation of a random variable as $\Expect\left\{S\right\}$ and the imaginary unit as $j$.

\emph{Sections}: Section II is dedicated to introduce the concept of spherical modulation and how the polarization can be used to transmit information. Section III describes the concept of digital polarizer, the sufficient conditions of the feasibility in the analog domain and the synthesis of digital polarizers. Section IV presents the framework to use the Cryptography to the physical realm by introducing the encryption and decryption algorithms, a fundamental analysis of the statistics and the introduction of two novel parameters: the Amount of Transformation and the Average Transformation. Section V introduces three different designs to perform the encryption of the electromagnetic fields: the Golden Encipherment, the Rotation Encipherment and the Opposite Encipherment. Section VI analyses analytically the impact of imperfect polarization in realistic environments, where cross polarization or unbalanced polarizations are present. Finally, Sections VII and VIII illustrate the results obtained by extensive and precise simulations and present the conclusions, respectively.

\section{MOTIVATION: SCENARIOS AND USE CASES}
The groundbreaking advancements in cryptographic systems at the physical layer (PHY) have opened up new possibilities for secure communication in various domains. The application of these systems extends beyond traditional encryption methods, offering unique advantages in specific scenarios and use cases. By harnessing the inherent properties of electromagnetic fields, particularly polarization, we can achieve robust and secure communication in challenging environments. 

\begin{enumerate}
	\item Military and Defense Applications: One of the most critical areas where these systems find immense value is in military and defense communications. The ability to establish secure and covert communication channels is of paramount importance in tactical operations, intelligence gathering, and strategic planning. By leveraging PHY-layer cryptographic systems, military personnel can transmit sensitive information without the risk of interception or compromise. The resilience of these systems against eavesdropping and their ability to maintain secure communication in hostile electromagnetic environments make them ideal for military applications.
\item Telecommunications and IoT Security: With the rapid growth of telecommunications networks and the Internet of Things (IoT), ensuring the security and integrity of data transmission has become a pressing concern. PHY-layer cryptographic systems offer a compelling solution for safeguarding sensitive information in these contexts. By integrating encryption and decryption processes directly into the communication channels, these systems can protect against unauthorized access, data breaches, and tampering. From secure voice and data transmission in telecommunication networks to securing the vast array of interconnected IoT devices, these systems provide a strong foundation for maintaining confidentiality and privacy.

\item Critical Infrastructure Protection: Critical infrastructure systems, including power grids, transportation networks, and water supply systems, are prime targets for malicious attacks. Ensuring the security and resilience of these infrastructures is crucial for national security and public safety. PHY-layer cryptographic systems can play a vital role in safeguarding critical infrastructure communications. By implementing robust encryption techniques at the PHY layer, the integrity and confidentiality of control signals and sensitive data can be maintained, mitigating the risk of cyber-attacks and unauthorized access.

\item Secure Sensor Networks: Sensor networks form the backbone of various applications, including environmental monitoring, surveillance systems, and industrial automation. These networks often operate in challenging and potentially hostile environments. PHY-layer cryptographic systems offer an added layer of security to protect the integrity and confidentiality of sensor data. By employing encryption techniques tailored to the specific characteristics of sensor networks, data can be securely transmitted, ensuring the accuracy and trustworthiness of the collected information.
\end{enumerate}

Physical layer security systems that leverage the polarization properties of electromagnetic fields offer a multitude of advantages over traditional cryptographic methods. By harnessing polarization as a key aspect of encryption and decryption processes, these systems provide unique benefits that enhance security and enable robust communication in various scenarios. The following are key advantages of utilizing physical layer security systems based on polarization:

\begin{enumerate}
	\item Inherent Security: Polarization-based encryption techniques offer inherent security by exploiting the fundamental characteristics of electromagnetic fields. Unlike traditional cryptographic methods that rely on complex algorithms and secret keys, polarization-based systems utilize the physical properties of light waves, making them resistant to algorithmic attacks and cryptanalysis. The inherent nature of polarization as a physical property ensures a strong foundation for secure communication.
\item Unpredictability and Randomness: The use of polarization for encryption introduces an element of unpredictability and randomness in the communication process. The polarization state of light can be manipulated and varied, generating an infinite number of possible encryption keys. This randomness adds an additional layer of complexity, making it extremely challenging for potential eavesdroppers to decipher the transmitted information. The inherent variability and unpredictability of polarization-based systems significantly enhance the security of the communication channels.

\item Robustness in Challenging Environments: Physical layer security systems based on polarization exhibit robustness in challenging and hostile electromagnetic environments. Polarization properties are relatively stable and resilient to external interference, such as electromagnetic noise and channel impairments. This robustness ensures that the encryption and decryption processes remain effective even in the presence of adverse conditions, making these systems suitable for applications in demanding environments, including military operations, critical infrastructure, and wireless communications.

\item Covert Communication: Leveraging polarization for encryption enables the possibility of covert communication. By exploiting specific polarization states as encryption keys, communication channels can be established that appear innocuous to potential eavesdroppers. The use of polarization-based techniques allows for stealthy transmission of sensitive information without raising suspicion or attracting unwanted attention. This covert communication capability is particularly valuable in military and defense scenarios, intelligence operations, and other contexts where secrecy and discretion are paramount.

\item Compatibility and Integration: Physical layer security systems based on polarization can be seamlessly integrated into existing communication infrastructures. These systems do not require significant modifications to the network infrastructure or the underlying communication protocols. They can be implemented at the PHY layer, enabling compatibility with various communication technologies, such as wireless networks, fiber-optic systems, and satellite communications. The ease of integration ensures that the benefits of polarization-based security can be realized without extensive infrastructure overhauls.
\end{enumerate}

\section{SPHERICAL MODULATION: THE FOUNDATION}
Before delving into the details of the proposed modulation technique, we introduce the concept of spherical modulation. Traditionally, the mapping of information bits onto electromagnetic propagation is accomplished using a two-dimensional plane, where the $x$ and $y$ axes correspond to the in-phase and quadrature (I/Q) components in the baseband model. This mathematical manipulation of the baseband model involves complex variable analysis, where complex numbers with real and imaginary parts are employed. This approach is made possible by the orthogonality of the in-phase and quadrature components of the electromagnetic fields.

However, in $1852$, Sir George Gabriel Stokes made a significant discovery regarding the polarization state of electromagnetic waves. He revealed that any polarized electromagnetic field, including completely unpolarized radiation like that emitted by the Sun, can be fully characterized by four parameters known as Stokes parameters \cite{Stokes1852}. These parameters provide a comprehensive description of the polarization state, capturing its intricacies.

Let consider a three-dimensional Cartesian basis (x, y and z axis). For any electromagnetic wave propagating along z-axis, the electric field $\vec{E}$ can be uniquely decomposed into two orthogonal components $E_x$ and $E_y$ with $(\hat{\mathbf{x}},\hat{\mathbf{y}})$ as the reference basis. Thus, if the angular frequency is denoted by $\omega$, it can be expressed as
\begin{equation}
\begin{split}
\vec{E}(z,t)&=\Re\left\{E_x(z,t)\hat{\mathbf{x}}+E_y(z,t)\hat{\mathbf{y}}\right\}\\
E_x(z,t)&=E_{0x}e^{j\left(\omega t-kz+\varphi_x\right)}=E_xe^{j\omega t-kz}\\
E_y(z,t)&=E_{0y}e^{j\left(\omega t-kz+\varphi_y\right)}=E_ye^{j\omega t-kz},
\end{split}
\end{equation}
where $E_{0x}$ and $E_{0y}$ are the amplitude of each component, $k$ is the wave number and $\varphi_x$ and $\varphi_y$ are their respective phases. The contribution of $\vec{E}_0$ can be decomposed by $E_{0x}$ and $E_{0y}$ as follows
\begin{equation}
\vec{E}_0=\begin{pmatrix}E_x\\E_y\end{pmatrix}=\begin{pmatrix}E_{0x}e^{j\varphi_x}\\E_{0y}e^{j\varphi_y}\end{pmatrix}.
\label{eq:jones}
\end{equation}
The vector $\vec{E}_0$ is called the Jones vector.

The four Stokes parameters describe the polarization state and are multiple defined as
\begin{equation}
\begin{split}
S_0&=|E_x|^2+|E_y|^2=E_xE_x^*+E_yE_y^*=E_{0x}^2+E_{0y}^2\\
S_1&=|E_x|^2-|E_y|^2=E_xE_x^*-E_yE_y^*=E_{0x}^2-E_{0y}^2\\
S_2&=2\Re\left\{E_xE_y^*\right\}=E_xE_y^*+E_x^*E_y=2E_{0x}E_{0y}\cos\theta\\
S_3&=-2\Im\left\{E_xE_y^*\right\}=j\left(E_xE_y^*-E_x^*E_y\right)=2E_{0x}E_{0y}\sin\theta,
\label{eq:j2s}
\end{split}
\end{equation}
where $\theta=\varphi_x-\varphi_y$.  

Stokes components are real quantities and characterize the state of the polarization for an arbitrary incident electromagnetic wave. The four Stokes parameters also meet the following inequality
\begin{equation}
S_0^2\geq S_1^2+S_2^2+S_3^2,
\label{eq:polineq}
\end{equation}
which is fulfilled with equality when the electromagnetic wave is completely polarized. 

It is important to note that the three parameters, $S_1$, $S_2$, and $S_3$, span a three-dimensional polarization space, allowing for the expression of all polarization states as linear combinations of these parameters. In the case of a fully polarized wave, the parameter $S_0$ can be derived from the other three parameters. As a result, the three parameters define a sphere centered at the origin with a radius of $S_0$, known as the Poincaré Sphere \cite{Poincare1892}.

This Poincaré Sphere serves as a convenient reference grid for constellation design. Instead of using conventional two-dimensional grids, we can employ a three-dimensional grid that aligns with the sphere's surface to define our constellation. It is important to emphasize that this three-dimensional space is an abstract mathematical representation that characterizes all possible polarization states. Therefore, each point within this space corresponds to a specific polarization state, with its own associated energy.

By leveraging this spherical representation, we gain a richer framework for designing constellations and modulating information. The three-dimensional grid provided by the Poincaré Sphere allows for more diverse and nuanced constellation points, offering increased capacity for encoding information within the polarization state of the electromagnetic wave. This approach harnesses the energy and potential of the polarization space, enabling more efficient and robust communication systems. 

The conversion from Stokes vector to Jones vector (the electric field) can be found in \cite{Henarejos2018}, which is described by
\begin{equation}
\vec{E}_0=
\begin{pmatrix}\sqrt{\frac{S_0+S_1}{2}}e^{-j\theta}\\\sqrt{\frac{S_0-S_1}{2}}e^{j\theta}\end{pmatrix},
\label{eq:stokes2E}
\end{equation}
where $\tan\theta=\frac{S_3}{S_2}$. In spherical coordinates it takes the form
\begin{equation}
	\vec{E}_0=
	\begin{pmatrix}\sqrt{\mathcal{E}}\cos\frac{\vartheta}{2}e^{-j\frac{\phi}{2}}\\\sqrt{\mathcal{E}}\sin\frac{\vartheta}{2} e^{j\frac{\phi}{2}}\end{pmatrix},
	\label{eq:stokes2Esph}
\end{equation}
where $\phi\in[0,2\pi)$, $\vartheta\in[0,\pi)$ and $\mathcal{E}\in\mathbb{R}^+$ are the azimuthal, elevation and radii components of the sphere, respectively.

Designing an efficient constellation on a sphere involves equally distributing points to maximize the minimum distance, a problem known as the Tammes problem \cite{Tammes1930}. Researchers have addressed this challenge using mathematical optimization techniques, with works by Hardin and Sloane \cite{Hardin1995, Sloane1998} and Conway \cite{Conway2013} proposing solutions for specific values of points. Mapping bits to symbols on the sphere's surface is addressed in Henarejos et al.'s work \cite{Henarejos2018}, providing tables for constellation sizes of $2$, $4$, $8$, and $16$ points. These contributions enhance the efficiency and capacity of communication systems utilizing spherical modulation.

\section{DIGITAL POLARIZERS}
A polarizer is an object capable of modifying the polarization of an incident electromagnetic wave. In the field of optics, polarizers are commonly used and are made of materials with unique properties that alter the conditions of reflected and refracted light. For example, polarized sunglasses block incoming sunlight except for a specific slanted polarization that is allowed to pass through.

Various types of polarizers exist, including linear polarizers, retarders, and rotators, each capable of modifying the incident polarization to produce a refracted or reflected wave with a different polarization. This transformation can be described by the expression:
\begin{equation}
	\vec{E}_0' = \mathbf{J}\vec{E}_0,
\end{equation}
where $\vec{E}_0'$ represents the refracted or reflected Jones vector, and $\mathbf{J}$ is the polarizer's matrix known as the Jones matrix, which governs the transformation of the incident polarization.

While there are various types of polarizers, it is more convenient to use the formulation introduced by Mueller. This formulation converts the $2\times 2$ Jones matrix into a $4\times 4$ matrix known as the Mueller matrix, as proposed by Heinrich Mueller. Hence, every Jones matrix has the equivalent Mueller matrix, which is described by
\begin{equation}
	\matrix{M}=\matrix{A}\left(\matrix{J}\otimes\matrix{J}^*\right)\matrix{A}^{-1},
	\label{eq:mueller_orig}
\end{equation}
where $\matrix{J}^*$ is the conjugate Jones matrix ($\matrix{J}^*\equiv \left[J_{nm}^*\right]$) and $\matrix{A}$ is a unitary matrix that takes the form
\begin{align}
	\matrix{A}=\begin{pmatrix}
		1 & 0 & 0 & 1\\
		1 & 0 & 0 & -1\\
		0 & 1 & 1 & 0\\
		0 & j & -j & 0
	\end{pmatrix}, &\ 
	\matrix{A}^{-1}=\frac{1}{2}\matrix{A}^H.
\end{align}

Before continue, we introduce the following definition:
\begin{definition}
	Let $\mathcal{J}\subset\mathbb{C}^{2\times 2}$ be the set of Jones matrices. Let $\mathcal{M}\subset\mathbb{R}^{4\times 4}$ be the set of Mueller matrices. Let $\xi$ be the injective function that maps Jones matrices to Mueller matrices. Then we can find a subset of Mueller matrices $\mathcal{M}^P\subset\mathcal{M}$ that is uniquely generated from the set of Jones matrices:
	\begin{align*}
		\xi \colon\mathcal{J} &\to \mathcal{M}^P\\
		\matrix{J} &\mapsto \matrix{A}\left(\matrix{J}\otimes\matrix{J}^*\right)\matrix{A}^{-1}.
	\end{align*}
	\label{def:jonesmueller}
\end{definition}

Definition \ref{def:jonesmueller} describes that every Jones matrix has an equivalent Mueller matrix but this premise is not reciprocally: not every Mueller matrix has an equivalent Jones matrix. Therefore, we can ensure that if $\matrix{M}\in\mathcal{M}^P$, then it is physically reproducible.

From \eqref{eq:mueller_orig} we can state the following lemma:
\begin{lemma}
	Given the set of Mueller matrices $\mathcal{M}^P$ generated from the set of Jones matrices, for every Mueller matrix, its trace and determinant are non-negative:
	\begin{equation}
		\begin{array}{l}
			\tr\left(\matrix{M}^k\right)\geq 0\\
			\det\left(\matrix{M}^k\right)\geq 0
		\end{array},\ \forall \matrix{M}\in\mathcal{M},\ \forall k \geq 0.
	\end{equation}
\label{lemma:mueller_trace}
\end{lemma}
\begin{proof}
	To proof the lemma we apply the trace operator to \eqref{eq:mueller_orig}. Hence,
	\begin{subequations}
		\begin{align}
		\tr\left(\matrix{M}^k\right)&=\tr\left(\matrix{A}\left(\matrix{J}\otimes\matrix{J}^*\right)\matrix{A}^{-1}\ldots\matrix{A}\left(\matrix{J}\otimes\matrix{J}^*\right)\matrix{A}^{-1}\right)\nonumber\\
			&=\tr\left(\left(\matrix{J}\otimes\matrix{J}^*\right)^k\right)\nonumber\\
			&=\tr\left(\matrix{J}^k\otimes\matrix{J}^{k*}\right)\nonumber\\
			&=\tr\left(\matrix{J}^k\right)\tr\left(\matrix{J}^k\right)^*\nonumber\\
			&=\left|\tr\left(\matrix{J}^k\right)\right|^2\label{eq:trMtrJ}\\
			&\geq 0\nonumber\\
			\det\left(\matrix{M}^k\right)&=\det\left(\matrix{J}^k\otimes\matrix{J}^{k*}\right)\nonumber\\
			&=\det\left(\matrix{J}^k\right)^2\det\left(\matrix{J}^k\right)^{2*}\nonumber\\
			&=\left|\det\left(\matrix{J}\right)^k\right|^4\label{eq:detMdetJ}\\
			&\geq 0.\nonumber
		\end{align}
	\end{subequations}
\end{proof}
Note that Lemma \ref{lemma:mueller_trace} does not make any assumption on the sign of eigenvalues. Expressions \eqref{eq:trMtrJ} and \eqref{eq:detMdetJ} are interesting since both express the trace and determinant operators of Mueller matrix as a function of Jones matrix. Specially, with \eqref{eq:detMdetJ} and the relationships introduced in \cite{Xing1992}, we propose the following lemma:
\begin{lemma}
	Given the set of Mueller matrices $\mathcal{M}^P$ generated from the set of Jones matrices, for every Mueller matrix, its determinant can be expressed as a function of a single row or a single column as follows:
	\begin{equation}
			\begin{split}
		\det\left(\matrix{M}\right)&=\left(M_{00}^2-M_{01}^2-M_{02}^2-M_{03}^2\right)^2\\
		&=\left(M_{10}^2-M_{11}^2-M_{12}^2-M_{13}^2\right)^2\\
		&=\left(M_{20}^2-M_{21}^2-M_{22}^2-M_{23}^2\right)^2\\
		&=\left(M_{30}^2-M_{31}^2-M_{32}^2-M_{33}^2\right)^2\\
		&=\left(M_{00}^2-M_{10}^2-M_{20}^2-M_{30}^2\right)^2\\
		&=\left(M_{01}^2-M_{11}^2-M_{21}^2-M_{31}^2\right)^2\\
		&=\left(M_{02}^2-M_{12}^2-M_{22}^2-M_{32}^2\right)^2\\
		&=\left(M_{03}^2-M_{13}^2-M_{23}^2-M_{33}^2\right)^2.\\
		\end{split}
	\end{equation}
\label{lemma:det_mueller}
\end{lemma}

The Mueller matrix transforms the Stokes vector, as the Jones matrix transforms the Jones vector. Thus,
\begin{equation}
	\vec{S}'=\matrix{M}\vec{S},
\end{equation}
where $\vec{S}$ is the Stokes vector of the incident electromagnetic wave, $\matrix{M}\in\mathbb{R}^{4\times 4}$ is the Mueller matrix, and $\vec{S}'$ is the Stokes vector of the refracted/reflected wave. The Stokes transformation can be seen as a transformation of the four-dimensional vector and, therefore, as a transformation of the Poincaré Sphere. 

\subsection{SUFFICIENT CONDITIONS FOR PHYSICALLY REALIZABLE MUELLER MATRIX}
As elucidated, the implementation of spherical modulation necessitates the modulation of information onto the Stokes vector, which serves as an orthogonal basis for the Poincaré sphere. In addition, the utilization of the Stokes-Mueller calculus offers notable advantages, as it encompasses the characterization of all polarized waveforms, including those that are partially or entirely unpolarized. However, while every Jones matrix possesses an equivalent Mueller matrix, the converse is not universally applicable. Therefore, although we possess the flexibility to manipulate the Stokes vector in an arbitrary manner, it is imperative that a valid translation between the Stokes and Jones vectors exists.

In the ensuing discussion, we introduce the concept of \emph{pure Mueller matrices}, denoting a set of matrices that characterize realizable systems which are deterministic, non-overpolarizing, and non-depolarizing. These matrices guarantee that the equivalent system neither introduces additional energy nor depolarizes the incident wave. Furthermore, we establish the notion of \emph{golden Mueller matrices} as a subset of pure Mueller matrices that consistently yield refracted or reflected waveforms entirely polarized, thereby possessing an equivalent Jones matrix.

Despite the presence of 16 degrees of freedom associated with $4 \times 4$ matrices, it is important to acknowledge that not all matrices within this framework are physically realizable or qualify as pure Mueller matrices according to extant literature \cite{Gil2000}. Indeed, to ensure the convertibility of the transformed Stokes vector $\vec{S}'$ into the Jones vector, as expressed by equations (\ref{eq:stokes2E}) and (\ref{eq:stokes2Esph}), it is imperative that the Mueller matrix satisfies the criteria of being a pure Mueller matrix. Thus, a Mueller matrix is a physical Mueller matrix if it meets the following conditions:
\begin{enumerate}
	\item Eigenvalues condition (pure Mueller matrix): the coherency matrix $\matrix{C}$ is positive semi-definite (non-negative eigenvalues):
	\begin{align}
		\lambda_0 \geq \lambda_1 \geq \lambda_2 \geq \lambda_3 \geq 0,
		\label{eq:cond1}
	\end{align}
    where $\lambda_i, i\in\{0,1,2,3\}$ are the eigenvalues of $\matrix{C}$. It is worth to mention that the eigenvalue condition is referred to the coherency matrix. By Lemma \ref{lemma:mueller_trace}, the relative Mueller matrix is not constrained to the positivity of its eigenvalues.
	\item Transmittance condition: 
	\begin{subequations}
		\begin{align}
			g_f&=M_{00}+\sqrt{M_{01}^2+M_{02}^2+M_{03}^2}&\leq 1\label{eq:cond21}\\
			g_r&=M_{00}+\sqrt{M_{10}^2+M_{20}^2+M_{30}^2}&\leq 1\label{eq:cond22},
		\end{align}	
	\end{subequations}
\end{enumerate}

The coherency matrix is an Hermitian matrix ($\matrix{C}=\matrix{C}^H$), which is defined as 
\begin{subequations}
	\begin{align}
		\matrix{C}&=\frac{1}{4}\sum_{n,m}M_{nm}\matrix{\Gamma}_{nm}\\
		\matrix{\Gamma}_{nm}&=\matrix{A}\left(\matrix{\sigma}_n\otimes\matrix{\sigma}_m^*\right)\matrix{A}^{-1},
		\label{eq:gammadef}
	\end{align}
\end{subequations}
and it fully characterizes the Mueller matrix. Furthermore, the Mueller matrix can be expressed as a function of the coherency matrix as
\begin{equation}
	M_{nm}=\tr\left(\matrix{\Gamma}_{nm}\matrix{C}\right).
	\label{eq:coherency2mueller}
\end{equation}

Additionally, $\matrix{\sigma}_n$ are the Pauli matrices and they are defined as
\begin{equation}
	\begin{split}
	\matrix{\sigma}_0=\begin{pmatrix} 1 & 0 \\ 0 & 1 \end{pmatrix},\qquad & \matrix{\sigma}_1=\begin{pmatrix} 1 & 0 \\ 0 & -1 \end{pmatrix} \\
	\matrix{\sigma}_2=\begin{pmatrix} 0 & 1 \\ 1 & 0 \end{pmatrix},\qquad & \matrix{\sigma}_3=\begin{pmatrix} 0 & -j \\ j & 0 \end{pmatrix}.
	\end{split}
\end{equation}

The coherency matrix is specially interesting due to its Hermitian symmetry. Pure Mueller matrices are those that do not modify the polarizing degree and the emerging waveform is totally polarized. Hence, the following inequality holds
\begin{multline}
	\frac{1}{4}\tr\left(\matrix{M}^T\matrix{M}\right)=\tr\left(\matrix{C}^2\right)=\sum_id_i\lambda_i^2\\
	\leq\left(\sum_id_i\lambda_i\right)^2=\tr\left(\matrix{C}\right)^2=M_{00}^2,
	\label{eq:ineqeigen}
\end{multline}
where $\lambda_i$ are the four eigenvalues of $\matrix{C}$ and is equally fulfilled for pure Mueller matrices \cite{Gil2000}. In that case, when the equal sign is fulfilled for pure Mueller matrices, there is only a single eigenvalue strictly positive ($\lambda_0>0,\ \lambda_1=\lambda_2=\lambda_3=0$).

The coherence matrix is often considered more useful than the Mueller matrix in certain applications and scenarios due to its ability to capture the statistical properties of partially polarized light. While the Mueller matrix provides a deterministic description of the transformation of polarization states, the coherence matrix offers a statistical representation that incorporates information about the correlations between different polarization components.

Here are a few reasons why the coherence matrix is often preferred over the Mueller matrix in certain contexts:
\begin{enumerate}
	\item Statistical Description: The coherence matrix provides a statistical characterization of the polarization properties of light. It captures not only the transformation of polarization states but also the correlations and fluctuations in the polarization components. This is particularly useful when dealing with partially polarized light sources or when analyzing the impact of environmental factors on polarization, such as scattering or depolarization.
	\item Degree of Polarization: The coherence matrix allows for the direct calculation of polarization parameters, such as the degree of polarization. This parameter quantifies the extent to which the light is polarized and provides insights into the quality of polarization in a given system. It is a valuable metric in applications where polarimetric information is important, such as remote sensing or polarization imaging.
	\item Polarization Statistics: By analyzing the eigenvalues and eigenvectors of the coherence matrix, one can extract valuable information about the polarization statistics, including the orientation of the principal polarization states and the polarization purity. These parameters offer a deeper understanding of the polarization characteristics of the light and can be used to optimize polarization-based systems or identify specific polarization effects.
	\item Partially Polarized Light: The coherence matrix is particularly well-suited for dealing with partially polarized light, which is commonly encountered in many practical scenarios. It provides a comprehensive representation of the polarization properties, even in cases where the light exhibits varying degrees of polarization or undergoes fluctuations over time or space.
\end{enumerate}

While the Mueller matrix remains valuable for deterministic polarization transformations and analyzing fully polarized light, the coherence matrix offers a more comprehensive and statistical perspective on polarization. Its use is especially prevalent in fields such as polarimetry, optical communication, remote sensing, and biomedical optics, where the analysis of partially polarized light and its statistical properties is of paramount importance.

\subsection{SYNTHESIS OF DIGITAL POLARIZERS}
The preceding section outlined the essential conditions for designing a realizable Mueller matrix capable of manipulating incident polarization. In this section, we delve into the synthesis of digital polarizers capable of generating arbitrarily polarized electromagnetic waveforms.

As stated in Equation \eqref{eq:jones}, the Jones vector represents the electric field of the system and can be processed digitally through sampling, discretization, and subsequent manipulation. The Jones vector utilizes a two-dimensional orthonormal basis, typically represented by $\mathbf{\hat{x}}$ and $\mathbf{\hat{y}}$, which can be realized using two orthogonal and precisely calibrated dipoles. By employing these dipoles, we can radiate an electromagnetic waveform encompassing all possible polarizations defined by the Poincaré Sphere. The reciprocity theorem ensures the existence of a corresponding system at the receiver's end.

Additionally, due to the Hermitian symmetry of the coherency matrix $\matrix{C}$, designing the digital polarizer becomes less complex by utilizing the coherency matrix rather than directly designing the Mueller matrix. Through discretizing the Jones vector, we can synthesize a \emph{digital polarizer} by designing an artificial Mueller matrix to transform the original Jones vector into another Jones vector. The overall process can be summarized as follows:

\begin{enumerate}
	\item Design a coherency matrix that satisfies the conditions of eigenvalues and transmittance.
	\item Compute the Stokes vector corresponding to the discrete Jones vector, which represents a digital signal prior to digital-to-analog conversion (DAC).
	\item Transform the Stokes vector using the precomputed Mueller matrix.
	\item Obtain the transformed Jones vector by applying Equation \eqref{eq:stokes2E}.
	\item Pass the transformed Jones vector to the DAC.
\end{enumerate}

Thus, with a calibrated dual-polarized antenna, it becomes possible to transmit all polarizations using a predefined pattern characterized by the Mueller matrix. At the receiver's end, a dual-polarized antenna must also be employed to recover the transmitted signal. The crucial aspect, further explored in the next section, lies in the fact that without knowledge of either the Jones or Mueller matrices, the receiver cannot recover the original Jones vector employed by the transmitter.

\section{POLARIZATION FOR PHYSICAL LAYER SECURITY}
Encryption in cryptography aims to obfuscate information, referred to as plaintext. Unlike traditional cryptographic encryption, the focus shifts to securing the physical representation of the data. Instead of using secret keys and encryption algorithms, the concept of physical layer security involves manipulating the electromagnetic field to obfuscate the original signal.

In our proposed system, we introduce the notion of a \emph{plainfield}, which represents the original electromagnetic field, and a \emph{cipherfield}, which corresponds to the obfuscated electromagnetic field. The goal is to transform the plainfield into the cipherfield in a manner that cannot be deciphered by an unauthorized party without knowledge of the secret key.

To implement this obfuscation/deobfuscation system, we utilize the sphere constellation introduced in the previous section. The information bits are mapped onto symbols on the surface of the Poincaré sphere, resulting in the modulated vector, denoted as $\vec{S}$.

Furthermore, we design and synthesize multiple Mueller matrices based on the secret pattern. It is crucial to note that the obfuscation process operates on a per-block basis, meaning that information is obfuscated in segments, and the internal states are reset for each block. This approach ensures the security of the transmitted data while maintaining efficiency and effectiveness.

We describe the entire system as follows
\begin{equation}
	\vec{E}_0'=\mathcal{S}^{-1}\left(\matrix{M}_{\vec{K}}\mathcal{C}\left(\vec{b}\right)\right),
	\label{eq:polciph}
\end{equation}
where $\vec{E}_0'$ is the \emph{cipherfield}, $\mathcal{S}^{-1}\left(\cdot\right)$ is the Stokes to Jones conversion defined in \eqref{eq:stokes2E}, $\matrix{M}_{\vec{K}}$ is the Mueller matrix generated from the secret pattern $\vec{K}$, $\mathcal{C}\left(\cdot\right)$ is the constellation mapping that maps the information bits $\vec{b}$ (plaintext) on the sphere modulation to obtain the Stokes vector, the \emph{plainfield}. Algorithm \ref{alg:enc} describes the pseudocode for the obfuscation process.

\begin{algorithm}
	\caption{Obfuscation}
	\begin{algorithmic}[1]
		\Require L (number of constellation symbols)
		\Require $\vec{b}=\begin{pmatrix}b_0 & \ldots & b_{B-1}\end{pmatrix}$ (input bits, $B=\lfloor\log_2 L\rfloor$)
		\Require $\matrix{M}_{\vec{K}}$
		\Ensure $\vec{E}'$ (output Jones vector)
		\State Map bits on the constellation: $s\leftarrow \mathcal{C}\left(\vec{b}\right)$.
		\State Obtain Stokes vector $\vec{S}\leftarrow s$.
		\State Encrypt $\vec{S}'\leftarrow\matrix{M}_{\vec{K}}\vec{S}$.
		\State Obtain $\vec{E}_0'$ by applying \eqref{eq:stokes2E}: $\vec{E}_0'\leftarrow\mathcal{S}^{-1}\left(\vec{S}'\right)$.
	\end{algorithmic}
	\label{alg:enc}
\end{algorithm}

In the same way, we define the system model in the equivalent baseband model as follows
\begin{equation}
	\vec{y}=\matrix{H}\vec{E}_0'+\vec{w},
\end{equation}
where $\vec{y}\in\mathbb{C}^2$ is the received field in two orthogonal polarizations, $\matrix{H}\in\mathbb{C}^{2\times 2}$ is the channel matrix which models the environment impairments and $\vec{w}\in\mathbb{C}^2$ is the Additive White Gaussian Noise (AWGN), which follows a complex normal distribution $\vec{w}\sim\mathcal{CN}\left(\vec{0},\sigma_w^2\matrix{I}\right)$. The decryption process is described by Algorithm \ref{alg:dec}, where the equalization step is denoted by $\mathcal{H}_{\matrix{H}}$.

\begin{algorithm}
	\caption{Deobfuscation}
	\begin{algorithmic}[1]
		\Require $\matrix{H}$
		\Require $\vec{y}$
		\Require $\matrix{M}_{\vec{K}}$
		\Ensure $\vec{b}$
		\State Equalize the received signal $\vec{\tilde{y}}\leftarrow \mathcal{H}_{\matrix{H}}\left(\vec{y}\right)$.
		\State Obtain Stokes vector $\vec{\tilde{S}}\leftarrow \mathcal{S}\left(\vec{\tilde{y}}\right)$.
		\State Obtain the original Stokes vector (\emph{plainfield}) $\vec{S}\leftarrow\matrix{M}_{\vec{K}}^{-1}\vec{\tilde{S}}$.
		\State Obtain the bits (plaintext) by demapping $\vec{b}\leftarrow\mathcal{C}^{-1}\left(\vec{S}\right)$.
	\end{algorithmic}
	\label{alg:dec}
\end{algorithm}

From \eqref{eq:polciph} we can observe that $\matrix{M}_{\vec{K}}$ plays a fundamental role and it determines the robustness of the system. Furthermore, we introduced the notation of $\matrix{M}_{\vec{K}}^{-1}$, which corresponds to the inverse of the Mueller matrix used during the obfuscation. Hence, we introduce a third condition to design an encryption system:
\begin{enumerate}
	\setcounter{enumi}{2}
	\item The Mueller matrix $\matrix{M}$ must be invertible. This is equivalent to ensure the determinant is non-zero,  or, equivalently, one of the following inequalities:
	\begin{equation}
		\begin{array}{ll}
			&M_{0i}^2\neq M_{1i}^2+M_{2i}^2+M_{3i}^2\\
			&M_{i0}^2\neq M_{i1}^2+M_{i2}^2+M_{i3}^2
		\end{array},\ \forall i\in[0,1,2,3].
	\label{eq:cond3}
	\end{equation}
\end{enumerate}

\subsection{STOKES VECTOR STATISTICS}
\label{sect:stokes_stats}
In the step $2$ of Algorithm \ref{alg:dec} the received signal $\vec{\tilde{y}}$ is converted to the received Stokes vector. This is a non-linear process which modifies the noise. In this section we analyze the impact of the noise contribution, by assuming an ideal and unitary channel equalization that does not alter the noise distribution. Recalling that 
\begin{equation}
	\vec{\bar{y}}=\begin{pmatrix}y_1\\y_2\end{pmatrix}=\begin{pmatrix}x_1\\x_2\end{pmatrix}+\begin{pmatrix}w_1\\w_2\end{pmatrix},
\end{equation}
where $x_1\equiv E_x$ and $x_2\equiv E_y$, the Stokes vector at reception can be reformulated as follows
\begin{equation}
	\begin{split}
		\tilde{S}_0&=|y_1|^2+|y_2|^2=|x_1+w_1|^2+|x_2+w_2|^2\\
		\tilde{S}_1&=|y_1|^2-|y_2|^2=|x_1+w_1|^2-|x_2+w_2|^2\\
		\tilde{S}_2&=2\Re\left\{y_1y_2^*\right\}=2\Re\left\{x_1x_2^*+x_1w_2^*+w_1x_2^*+w_1w_2^*\right\}\\
		\tilde{S}_3&=-2\Im\left\{y_1y_2^*\right\}=-2\Im\left\{x_1x_2^*+x_1w_2^*+w_1x_2^*+w_1w_2^*\right\}.
	\end{split}
\end{equation}

The variance of the Stokes vector is expressed as
\begin{equation}
	\sigma_{\vec{\tilde{S}}}^2=\Expect\left\{\vec{\tilde{S}}^2\right\}-\Expect\left\{\vec{\tilde{S}}\right\}^2.
	\label{eq:sigma_s}
\end{equation}

For the sake of clarity, we introduce the notation $\Re\left\{x\right\}=x^R$ and $\Im\left\{x\right\}=x^I$. We assume that $\vec{x}$ is zero mean, uncorrelated and the power is equally distributed in $x_1$ and $x_2$, where $\Expect\left\{|x_1|^2\right\}=\Expect\left\{|x_2|^2\right\}=P_x$. Note that if $a,b\in\mathbb{C}$ are uncorrelated
\begin{equation}
	\begin{split}
	\Expect\left\{\Re\left\{ab^*\right\}\right\}&=\Expect\left\{a^R\right\}\Expect\left\{b^R\right\}+\Expect\left\{a^I\right\}\Expect\left\{b^I\right\}\\
	\Expect\left\{\Im\left\{ab^*\right\}\right\}&=\Expect\left\{a^I\right\}\Expect\left\{b^R\right\}-\Expect\left\{a^R\right\}\Expect\left\{b^I\right\}
\end{split}
\end{equation}
 The expectation of the Stokes vector is expressed by
\begin{equation}
	\begin{split}
		\Expect\left\{\tilde{S}_0\right\}&=\Expect\left\{\|\vec{x}\|^2\right\}+\Expect\left\{\|\vec{w}\|^2\right\}=2\left(P_x+\sigma_w^2\right)\\
		\Expect\left\{\tilde{S}_1\right\}&=\Expect\left\{|x_1|^2-|x_2|^2\right\}+\Expect\left\{|w_1|^2-|w_2|^2\right\}=0\\
		\Expect\left\{\tilde{S}_2\right\}&=2\Expect\left\{\Re\left\{x_1x_2^*+x_1w_2^*+w_1x_2^*+w_1w_2^*\right\}\right\}=0\\
		\Expect\left\{\tilde{S}_3\right\}&=-2\Expect\left\{\Im\left\{x_1x_2^*+x_1w_2^*+w_1x_2^*+w_1w_2^*\right\}\right\}=0\\
	\end{split}
\label{eq:avgs}
\end{equation}
and thus
\begin{equation}
	\vec{\bar{\tilde{S}}}=\Expect\left\{\vec{\tilde{S}}\right\}=2\begin{pmatrix}P_x+\sigma_w^2\\0\\0\\0\end{pmatrix}.
\end{equation}
This makes sense, since $S_0$ measures the incident energy of both polarizations and, therefore, is a non-zero mean random variable distributed in the domain $[0,+\infty)$. $S_1$, $S_2$ and $S_3$ are random variables corresponding to the axis of the sphere constellation, centred at the origin and, thus, all are zero mean.

After some mathematical manipulations, and recalling that the kurtosis of the Gaussian random variable is $3\sigma_w^4/4$, the non-central second order expectations of Stokes parameters are expressed by
\begin{equation}
	\begin{split}
		\Expect\left\{\tilde{S}_0^2\right\}&=\Expect\left\{\left(\|\vec{x}\|^2+\|\vec{w}\|^2\right)^2\right\}\\
		&=4P_x^2+12P_x\sigma_w^2+6\sigma_w^4\\
		\Expect\left\{\tilde{S}_1^2\right\}&=\Expect\left\{\left(|x_1+w_1|^2-|x_2+w_2|^2\right)^2\right\}\\
		&=4P_x\sigma_w^2+2\sigma_w^4\\
		\Expect\left\{\tilde{S}_2^2\right\}&=\Expect\left\{\left(\left(x_1+w_1\right)\left(x_2^*+w_2^*\right)+\left(x_1^*+w_1^*\right)\left(x_2+w_2\right)\right)^2\right\}\\
		&=\left(P_x+\sigma_w^2\right)^2\\
		\Expect\left\{\tilde{S}_3^2\right\}&=\Expect\left\{\left(\left(x_1+w_1\right)\left(x_2^*+w_2^*\right)-\left(x_1^*+w_1^*\right)\left(x_2+w_2\right)\right)^2\right\}\\
		&=\left(P_x+\sigma_w^2\right)^2
	\end{split}
\label{eq:noncentrals}
\end{equation}

Therefore, by plugging \eqref{eq:avgs} and \eqref{eq:noncentrals} in \eqref{eq:sigma_s}, we obtain
\begin{equation}
	\begin{split}
		\sigma_{\tilde{S}_0}^2&=4P_x\sigma_w^2+2\sigma_w^4\\
		\sigma_{\tilde{S}_1}^2&=4P_x\sigma_w^2+2\sigma_w^4=\sigma_{\tilde{S}_0}^2\\
		\sigma_{\tilde{S}_2}^2&=2\left(P_x+\sigma_w^2\right)^2=\sigma_{\tilde{S}_0}^2+2P_x^2\\
		\sigma_{\tilde{S}_3}^2&=2\left(P_x+\sigma_w^2\right)^2=\sigma_{\tilde{S}_0}^2+2P_x^2
	\end{split}
	\label{eq:sigma_s_full}
\end{equation}

Significantly, upon careful analysis of equation \eqref{eq:sigma_s_full}, it becomes evident that the nature of the noise is notably distinct. Instead of a straightforward additive process, the Stokes conversion exhibits an intrinsic squared detection mechanism, wherein both the signal and the noise are encompassed. As a result, a conjoined noise component, intricately intertwined with the signal, emerges, as denoted by the inclusion of the term $4P_x\sigma_w^2$ in each of the four Stokes parameters.

\subsection{SIGNAL TO NOISE RATIO OF STOKES VECTOR}
\label{sect:stokes_snr}
The non-linearity of the Jones-Stokes conversion process introduces distinct alterations to the Signal-to-Noise Ratio (SNR), which vary according to the specific Stokes parameter. By examining equation \eqref{eq:noncentrals}, we investigate the transformation of SNR by expanding the Stokes parameters, thereby enabling the identification of the signal and noise components involved.

\begin{align}
	\tilde{S}_0=&|x_1|^2+|x_2|^2&+2\Re\left\{x_1w_1^*+x_2w_2^*\right\}+|w_1|^2+|w_2|^2\nonumber\\
	\tilde{S}_1=&|x_1|^2-|x_2|^2&+2\Re\left\{x_1w_1^*-x_2w_2^*\right\}+|w_1|^2-|w_2|^2\nonumber\\		\tilde{S}_2=&2\Re\left\{x_1x_2^*\right\}&+2\Re\left\{x_1w_2^*+w_1x_2^*+w_1w_2^*\right\}\nonumber\\	\tilde{S}_3=&\underbrace{-2\Im\left\{x_1x_2^*\right\}}_{\text{Signal}}&-\underbrace{2\Im\left\{x_1w_2^*+w_1x_2^*+w_1w_2^*\right\}\qquad\ }_{\text{Noise}}.\nonumber\\
\end{align}

The SNR is computed as the ratio between the signal's power and noise's power. By denoting the input SNR as $\gamma=\frac{P_x}{\sigma_w^2}$, the SNR of each Stokes parameter is expressed as
\begin{equation}
	\begin{split}
		\text{SNR}_0&=\frac{4P_x^2}{4P_x\sigma_w^2+6\sigma_w^4}=\gamma\times\frac{1}{1+\frac{3}{2}\gamma^{-1}}\\
		\text{SNR}_1&=0\\
		\text{SNR}_2&=\frac{2P_x^2}{4P_x\sigma_w^2+2\sigma_w^4}=\gamma\times\frac{1}{2+\gamma^{-1}}\\				\text{SNR}_3&=\frac{2P_x^2}{4P_x\sigma_w^2+2\sigma_w^4}=\gamma\times\frac{1}{2+\gamma^{-1}}.\\	
		\label{eq:snr_gamma}
	\end{split}
\end{equation}

The transformation of the Signal-to-Noise Ratio (SNR) as described in \eqref{eq:snr_gamma} reveals interesting observations. Notably, the parameter $\text{SNR}_1$ consistently remains at zero regardless of the noise level. In the case of $S_0$, $S_2$, and $S_3$, the input SNR $\gamma$ undergoes a reduction by an inverse proportionality factor. Specifically, in the high SNR regime ($\gamma > 10$), the SNR for $S_0$ remains unchanged, which is to be expected as it measures the incident energy that remains constant at high SNR levels. Conversely, the SNR for $S_2$ and $S_3$ is halved in the high SNR regime, aligning with the fact that they capture the real and imaginary components of the metric, suggesting a division by two of the SNR. On the other hand, in the low SNR regime, $\text{SNR}_0$ is proportional to $\gamma^2$, while $\text{SNR}_2$ and $\text{SNR}_3$ are proportional to $\frac{2}{3}\gamma^2$.

In the subsequent sections, we will delve into the generation of $\matrix{M}_{\vec{K}}$, explore various strategies for designing $\matrix{M}_{\vec{K}}$, and discuss the process of encipherment, which can be either data flow-driven or block-driven in nature.

\subsection{SECRET PATTERN STRENGTH AND AMOUNT OF TRANSFORMATION}
Before delving into the design aspects of the secret pattern, it is essential to assess the strength of a security system. In order to determine the robustness of a secret pattern, we establish certain metrics that define what constitutes a strong pattern. Intuitively, a pattern is considered strong if the resulting constellation points, after obfuscation, are significantly displaced from their original symbols. Mathematically, the strength of the pattern is closely tied to the distance between the original points and the ciphered points, denoted as $\mathcal{D}=\|\vec{S}'-\vec{S}\|^2$.

Building upon this concept, we introduce a novel metric called the "amount of transformation," denoted as $\mathcal{Q}$, which is defined as
\begin{equation}
	\mathcal{Q}=\int_{\vec{S}}\mathcal{D}\diff\vec{S}.
	\label{eq:def_amount_transformation}
\end{equation}

The magnitude $\mathcal{Q}$ represents an absolute and positive value that quantifies the degree of transformation that a Mueller matrix can impose on each Stokes vector. It serves as a comparative measure to assess the strength of different patterns. A higher value of $\mathcal{Q}$ indicates a greater transformation of the Stokes vector. A value of $\mathcal{Q}=0$ implies that the matrix $\matrix{M}$ does not induce any transformation, as $\mathcal{D}=0$ for all $\vec{S}$. For example, this would be the case when $\matrix{M}$ is equal to the identity matrix ($\matrix{I}$).

The amount of transformation is a bounded metric, with the lower bound trivially set at 0. However, we present the following theorem that establishes tighter upper and lower bounds:
\begin{theorem}
	Let $\mathcal{Q}$ be the amount of transformation and $0$ a trivial lower bound. Given a deterministic Mueller matrix $\matrix{M}$ and $\|\matrix{A}\|^2_F=\tr\left(\matrix{A}\matrix{A}^H\right)$ the Frobenius norm of $\matrix{A}$, the amount of transformation is upper and lower bounded as follows:
	\begin{equation}
		\frac{4\pi}{3}\left\|\matrix{M}-\matrix{I}\right\|_F^2\leq\mathcal{Q}\leq 8\pi\left\|\matrix{M}-\matrix{I}\right\|_F^2.
	\end{equation}
\label{theorem:uplowbounds}
\end{theorem}

\begin{proof}
To proof this theorem we expand the definition of the amount of transformation $\mathcal{Q}$

\begin{equation}
	\mathcal{Q}=\int_{\vec{S}}\mathcal{D}\diff \vec{S}=\int_{\vec{S}}\left\|\left(\matrix{M}-\matrix{I}\right)\vec{S}\right\|^2\diff\vec{S}=\int_{\vec{S}}\vec{S}^H\matrix{D}\vec{S}\diff\vec{S},
\end{equation}
where $\matrix{D}=\left(\matrix{M}-\matrix{I}\right)^H\left(\matrix{M}-\matrix{I}\right)$. For the sake of generality, we rewrite the Stokes vector as
\begin{equation}
	\vec{S}=\begin{pmatrix}
		\|\vec{\bar{S}}\|\\\vec{\bar{S}}
	\end{pmatrix},
\end{equation}
where $\vec{\bar{S}}\in\mathbb{R}^3$ is the \emph{Reduced Stokes Vector} that spans the Poincaré sphere ($\vec{\bar{S}}=\begin{pmatrix}S_1 & S_2 & S_3\end{pmatrix}^T$).  Before computing the integral, we expand the integrand. By introducing the following definitions,
\begin{equation}
	\matrix{D}=\begin{pmatrix} D_{00} & \rvline&  & \vec{D}_h^H & \\\hline
		\vec{D}_v & \rvline&  & \matrix{D}_s &
	\end{pmatrix},
\end{equation}
the integrand can be expanded into
\begin{equation}
	\vec{S}^H\matrix{D}\vec{S}=\|\vec{\bar{S}}\|^2D_{00}+\|\vec{\bar{S}}\|\left(\vec{D}_h^H+\vec{D}_v^H\right)\vec{\bar{S}}+\vec{\bar{S}}^H\matrix{D}_s\vec{\bar{S}}.
\end{equation}
Therefore, the integral can be rewritten into
\begin{multline}
	\int_{\vec{S}}\vec{S}^H\matrix{D}\vec{S}\diff\vec{S}\\=\int_{\vec{\bar{S}}}\left(\|\vec{\bar{S}}\|^2D_{00}+\|\vec{\bar{S}}\|\left(\vec{D}_h^H+\vec{D}_v^H\right)\vec{\bar{S}}+\vec{\bar{S}}^H\matrix{D}_s\vec{\bar{S}}\right)\diff\vec{\bar{S}}.
	\label{eq:integral_sds}
\end{multline}

The first integral is straightforward to compute, as it equals to the area of the sphere, since $\|\vec{\bar{S}}\|^2$ is constant:
\begin{equation}
	\int_{\vec{\bar{S}}}\|\vec{\bar{S}}\|^2D_{00}\diff\vec{\bar{S}}=4\pi\|\vec{\bar{S}}\|^2D_{00}.
\end{equation}

The second and the third can be solved by introducing a change of variable, from Cartesian to spherical coordinates, in such a way that $\vec{\bar{S}}=\left(S_1,S_2,S_3\right)=\left(\sin\theta\cos\phi,\sin\theta\sin\phi,\cos\theta\right)$ and $\diff\vec{\bar{S}}=\sin\theta\diff\theta\diff\phi$. Hence,
\begin{multline}
	\int_{\vec{\bar{S}}}\|\vec{\bar{S}}\|\left(\vec{D}_h^H+\vec{D}_v^H\right)\vec{\bar{S}}\diff\vec{\bar{S}}\\
	=\|\vec{\bar{S}}\|	\int_{\vec{\bar{S}}}\left(D_{hv1}\sin\theta\cos\phi+D_{hv2}\sin\theta\sin\phi\right.\\\left.+D_{hv3}\cos\theta\right)\sin\theta\diff\theta\diff\phi=0,
\end{multline}
where $\vec{D}_{hv}=\vec{D}_v+\vec{D}_h$.

Finally, the third integral can be solved by expanding the product. After some mathematical manipulation, we obtain
\begin{multline}
	\vec{\bar{S}}^H\matrix{D}_s\vec{\bar{S}}\\=D_{11}S_1^2+D_{22}S_2^2+D_{33}S_3^2+\left(D_{12}+D_{21}\right)S_1S_2\\+\left(D_{13}+D_{31}\right)S_1S_3+\left(D_{23}+D_{32}\right)S_2S_3,
\end{multline}
where
\begin{equation}
	\matrix{D}_s=\begin{pmatrix}
		D_{11} & D_{12} & D_{13}\\D_{12} & D_{22} & D_{23}\\D_{31} & D_{32} & D_{33}.
	\end{pmatrix}
\end{equation}

The solutions of these six integrals are the following:
\begin{equation}
	\begin{split}
		\int_{\vec{\bar{S}}}D_{11}S_1^2\diff\vec{\bar{S}}&=D_{11}\int_0^{2\pi}\int_{0}^{\pi}\sin^3\theta\cos^2\phi\diff\theta\diff\phi=\frac{4\pi}{3}D_{11}\\
		\int_{\vec{\bar{S}}}D_{22}S_2^2\diff\vec{\bar{S}}&=D_{22}\int_0^{2\pi}\int_{0}^{\pi}\sin^3\theta\sin^2\phi\diff\theta\diff\phi=\frac{4\pi}{3}D_{22}\\
		\int_{\vec{\bar{S}}}D_{33}S_2^2\diff\vec{\bar{S}}&=D_{33}\int_0^{2\pi}\int_{0}^{\pi}\cos^2\theta\sin\theta\diff\theta\diff\phi=\frac{4\pi}{3}D_{33}\\
		&\int_{\vec{\bar{S}}}\left(D_{12}+D_{21}\right)S_1S_2\diff\vec{\bar{S}}\\
		&=\left(D_{12}+D_{21}\right)\int_0^{2\pi}\int_{0}^{\pi}\sin^3\theta\sin\phi\cos\phi\diff\theta\diff\phi\\
		&=0\\
		&\int_{\vec{\bar{S}}}\left(D_{13}+D_{31}\right)S_1S_3\diff\vec{\bar{S}}\\
		&=\left(D_{13}+D_{31}\right)\int_0^{2\pi}\int_{0}^{\pi}\sin^2\theta\cos\theta\cos\phi\diff\theta\diff\phi\\
		&=0\\
		&\int_{\vec{\bar{S}}}\left(D_{23}+D_{32}\right)S_2S_3\diff\vec{\bar{S}}\\
		&=\left(D_{23}+D_{32}\right)\int_0^{2\pi}\int_{0}^{\pi}\sin^2\theta\cos\theta\sin\phi\diff\theta\diff\phi\\
		&=0.
	\end{split}
\end{equation}

Therefore, by assuming the normalization of the reduced Stokes vector $\vec{\bar{S}}$ ($\left\|\vec{\bar{S}}\right\|^2=1$), we can solve the integral \eqref{eq:integral_sds}
\begin{equation}
	\begin{split}
		\int_{\vec{S}}\vec{S}^H\matrix{D}\vec{S}\diff\vec{S}&=4\pi\left(\|\vec{\bar{S}}\|^2D_{00}+\frac{1}{3}\left(D_{11}+D_{22}+D_{33}\right)\right)\\
		&\geq\frac{4\pi}{3}\tr\left(\matrix{D}\right),
	\end{split}
	\label{eq:int_sds_finalD}
\end{equation}
which equally holds when $D_{00}=0$.

By replacing the matrix $\matrix{D}$ by the Mueller matrix, $D_{ii}$ coefficients can be denoted as
\begin{equation}
	\begin{split}
		D_{00}&=\left(M_{00}-1\right)^2+M_{10}^2+M_{20}^2+M_{30}^2\\
		D_{11}&=\left(M_{11}-1\right)^2+M_{01}^2+M_{21}^2+M_{31}^2\\
		D_{22}&=\left(M_{22}-1\right)^2+M_{02}^2+M_{12}^2+M_{32}^2\\
		D_{33}&=\left(M_{33}-1\right)^2+M_{03}^2+M_{13}^2+M_{23}^2.\\
	\end{split}
\end{equation}

After some mathematical manipulations, we obtain the following inequality
\begin{equation}
	\begin{split}
	\int_{\vec{S}}\vec{S}^H\matrix{D}\vec{S}\diff\vec{S}&\geq\frac{4\pi}{3}\left(\left\|\matrix{M}\right\|_F^2-2\tr\left(\matrix{M}\right)+4\right)\\
	&=\frac{4\pi}{3}\left\|\matrix{M}-\matrix{I}\right\|_F^2
	\end{split}
	\label{eq:low_bound_sds}
\end{equation}

By applying the same analysis, we present an upper bound of the amount of transformation $\mathcal{Q}$ by using the Cauchy-Schwartz inequality to the integrand:
\begin{equation}
\begin{split}
	\left\|\left(\matrix{M}-\matrix{I}\right)\vec{S}\right\|^2&\leq\left\|\matrix{M}-\matrix{I}\right\|_F^2\left\|\vec{S}\right\|^2.
\end{split}
\end{equation}

Thus,
\begin{equation}
	\begin{split}
	\mathcal{Q}=\int_{\vec{S}}\left\|\left(\matrix{M}-\matrix{I}\right)\vec{S}\right\|^2\diff\vec{S}&\leq\int_{\vec{S}}\left\|\matrix{M}-\matrix{I}\right\|_F^2\left\|\vec{S}\right\|^2\diff\vec{S}\\
	&=8\pi\left\|\matrix{M}-\matrix{I}\right\|_F^2
\end{split}
\end{equation}
with the assumption of normalized Stokes vector. 
\end{proof}

Both upper and lower bounds allow us to obtain a criteria to maximize the amount of transformation. We derive the following corollary:
\begin{corollary}
	A secret pattern $\vec{K}$ is considered strong and secure if and only if it generates a Mueller matrix $\matrix{M}_{\vec{K}}$ such that maximizes $\mathcal{Q}$. This is achieved by Mueller matrices such that
	\begin{equation}
		\tr\left(\matrix{M}_{\vec{K}}\right)=0.
	\end{equation}
	\label{theorem:maxdist}
\end{corollary}
\begin{proof}	
The expression \eqref{eq:low_bound_sds} represents a lower bound of the mathematical quantity $\mathcal{Q}$, which is dependent on the matrix $\matrix{M}$. In order to maximize the expression \eqref{eq:low_bound_sds}, it is necessary to maximize the Frobenius norm of the Mueller matrix while minimizing its trace. According to \eqref{eq:ineqeigen}, we can observe that the Frobenius norm of the Mueller matrix is bounded above by $4M_{00}^2$, which occurs when the matrix $\matrix{M}$ is a golden Mueller matrix (with $M_{00}=1$). Therefore, utilizing Lemma \ref{lemma:mueller_trace}, the maximum value is obtained when the trace of $\matrix{M}$, denoted as $\tr\left(\matrix{M}\right)$, equals zero, and $\matrix{M}$ itself is a golden Mueller matrix.
\end{proof}

An interesting consequence of \ref{theorem:maxdist} is that also maximizes the upper bound introduced in Theorem \ref{theorem:uplowbounds}:
\begin{corollary}
	The amount of transformation $\mathcal{Q}$ is a finite magnitude. For every Mueller matrix $\matrix{M}$, the following inequality holds:
	\begin{equation}
		\mathcal{Q}\leq64\pi.
	\end{equation}
\label{theorem:finupp}
\end{corollary}
\begin{proof}
	This establishes an unequivocal and finite upper bound, which remains unaffected by any specific Mueller matrix, as a direct outcome of Corollary \ref{theorem:maxdist}. In the case of all golden Mueller matrices, it holds that $\left\|\matrix{M}\right\|^2_F=4$, consequently yielding $\left\|\matrix{M}-\matrix{I}\right\|^2_F=8$ and $\tr\left(\matrix{M}\right)=0$. From this, it becomes evident that $\mathcal{Q}$ is bounded by the inequality $\mathcal{Q}\leq64\pi$.
\end{proof}

The noteworthy implication of Corollary \ref{theorem:finupp} is that due to the finite nature of the quantity, there exists a well-defined solution that maximizes it. Notably, this optimal solution is attained specifically through the utilization of golden Mueller matrices.

\subsection{AVERAGE AMOUNT OF TRANSFORMATION}
The Amount of Transformation, denoted as $\mathcal{Q}$, represents an intrinsic property that remains invariant regardless of the specific constellation. It characterizes the degree of transformation experienced by each point within the constellation. However, it is important to note that the transformation is not uniformly applied to all points in the constellation, indicating that each point undergoes a distinct level of transformation.

Within this section, we propose the introduction of a novel metric referred to as the \emph{Average Transformation}. This metric serves to quantify the extent of transformation experienced by individual points within the constellation. In the continuous scenario, the Average Transformation is defined as follows:
\begin{equation}
	\begin{split}
		\mathcal{P}&=\int_{\vec{S}}\mathcal{D}P_{\mathrm{S}}\left(\vec{S}\right)\diff\vec{S}\\
		&=\int_{\vec{S}}\vec{S}^H\matrix{D}\vec{S}P_{\mathrm{S}}\left(\vec{S}\right)\diff\vec{S},
	\end{split}
\end{equation}
and in the discrete case, it is equivalent to
\begin{equation}
	\begin{split}
		\mathcal{P}&=\sum_n^N\mathcal{D}P_{\mathrm{S}}\left(\vec{S}_n\right)\\
		&=\sum_n^N\vec{S}_n^H\matrix{D}\vec{S}_nP_{\mathrm{S}}\left(\vec{S}_n\right),
	\end{split}
\label{eq:discrete_avgQ}
\end{equation}
where $N$ is the total size of the constellation and $P_{\mathrm{S}}\left(\vec{S}\right)$ is the probability density function or probability distribution of the Stokes vector.

In the particular case where all symbols are equiprobable, ($P_{\mathrm{S}}\left(\vec{S}_n\right)=\frac{1}{N})$, \eqref{eq:discrete_avgQ} can be simplified to
\begin{equation}
	\begin{split}
		\mathcal{P}&=\frac{1}{N}\sum_n^N\vec{S}_n^H\matrix{D}\vec{S}_n\\
		&=\tr\left(\matrix{D}\matrix{\Sigma}\right),
	\end{split}
\end{equation}
where
\begin{equation}
	\matrix{\Sigma}=\frac{1}{N}\sum_n^N\vec{S}_n\vec{S}_n^H
\end{equation}
is the autocorrelation matrix of the constellation.

The particular case where all symbols are uncorrelated and belong to the unitary sphere, the autocorrelation matrix is diagonal and takes the form
\begin{equation}
	\matrix{\Sigma}=\begin{pmatrix}
		\|\vec{\bar{S}}\|^2&&&\\&1/3&&\\&&1/3&\\&&&1/3
	\end{pmatrix},
\end{equation}
where the Average Transformation equals to
\begin{equation}
	\mathcal{P}=\|\vec{\bar{S}}\|^2D_{00}+\frac{1}{3}\left(D_{11}+D_{22}+D_{33}\right).
	\label{eq:avgQ_uncorr}
\end{equation}

Upon examination of \eqref{eq:avgQ_uncorr}, it becomes evident that the solution remains independent of the specific constellation employed. Specifically, upon closer inspection of \eqref{eq:int_sds_finalD}, assuming an uncorrelated constellation, we can establish the following relationship:

\begin{equation}
	\mathcal{Q}=4\pi\mathcal{P}.
	\label{eq:Q4piP}
\end{equation}

Consequently, \eqref{eq:Q4piP} elucidates that the degree of encipherment is negligibly influenced by the arrangement of points within the constellation. The spatial distribution of the points does not hold substantial significance. Instead, the primary determinant of the encryption's strength resides in the design of the Mueller matrix. Notably, the use of a golden Mueller matrix results in maximum transformation and subsequently maximum obfuscation.

The significance of \eqref{eq:Q4piP} lies in the fact that lower and upper bounds for $\mathcal{Q}$ can be correspondingly applied to $\mathcal{P}$. Consequently, the following inequality holds:

\begin{equation}
	\frac{1}{3}\left\|\matrix{M}-\matrix{I}\right\|^2_F\leq\mathcal{P}\leq2\left\|\matrix{M}-\matrix{I}\right\|^2_F\leq16.
	\label{eq:P_bounds}
\end{equation}

Furthermore, the upper bound specified in \eqref{eq:P_bounds} aligns with the proof provided by Corollary \ref{theorem:finupp}, which establishes that for golden Mueller matrices, $\left\|\matrix{M}-\matrix{I}\right\|^2_F=8$. Remarkably, this value precisely corresponds to the upper bound in \eqref{eq:P_bounds}. As anticipated, the maximization of $\mathcal{Q}$ equates to the maximization of $\mathcal{P}$, and this optimal outcome is achieved through the utilization of golden Mueller matrices.

\section{SECRET PATTERN DESIGN}
As elucidated in the preceding section, polarization can be leveraged as a mechanism for information obfuscation through the implementation of polarized modulations that map binary data onto a spherical surface. This obfuscation process entails the utilization of a confidential pattern, which is subsequently encoded within a matrix known as the Mueller matrix. The Mueller matrix facilitates the linear combination of various Stokes parameters, culminating in the generation of a symbol that cannot be demodulated without access to the secret pattern.

%
%

In the next sections we introduce several pattern designs.

\subsection{GOLDEN ENCIPHERMENT}
\label{sect:golden}
In this section, we delve into the design of a golden Mueller matrix based on a secret pattern. To achieve this, the Mueller matrix must satisfy three conditions (\eqref{eq:cond1}, \eqref{eq:cond21}, \eqref{eq:cond22}) with the equality sign. Additionally, since pure Mueller matrices possess a single eigenvalue, we can express $M_{00}=\lambda$ and $\matrix{C}=\lambda\vec{c}\vec{c}^H$, where $\vec{c}\in\mathbb{C}^4$ and $\vec{c}^H\vec{c}=1$. By specializing equation \eqref{eq:coherency2mueller} for pure Mueller matrices, it can be reformulated as follows:
\begin{equation}
	M_{nm}=\tr\left(\matrix{\Gamma}_{nm}\matrix{C}\right)=\tr\left(\matrix{\Gamma}_{nm}\lambda\vec{c}\vec{c}^H\right)=\lambda\vec{c}^H\matrix{\Gamma}_{nm}\vec{c}.
	\label{eq:cohe2muepure}
\end{equation}

To guarantee that an arbitrary four-dimensional vector may generate golden Mueller matrix, we introduce the following theorem:
\begin{theorem}
A vector $\vec{c}\in\mathbb{C}^4$ only generates pure Mueller matrix if and only if can be expressed as 
\begin{equation}
	\begin{split}
	\vec{c}&=\begin{pmatrix}0\\k_1\\k_2\\k_3\end{pmatrix}e^{jk_0},\ \forall (k_0,k_1,k_2,k_3)\in\mathbb{R}^4\\
	\textrm{s.t.}\ \|\vec{c}\|^2&=1.
	\end{split}
	\label{eq:solution}
\end{equation}
\end{theorem}

\begin{proof}
To generate golden Mueller matrix, the coherency matrix has to meet the eigenvalue and transmittance conditions. To satisfy the first transmittance condition, we plug \eqref{eq:cohe2muepure} into \eqref{eq:cond21}
\begin{equation}
	\lambda+\sqrt{\left(\lambda\vec{c}^H\matrix{\Gamma}_{01}\vec{c}\right)^2+\left(\lambda\vec{c}^H\matrix{\Gamma}_{02}\vec{c}\right)^2+\left(\lambda\vec{c}^H\matrix{\Gamma}_{03}\vec{c}\right)^2}=1,
\end{equation}
which can be simplified to
\begin{equation}
		\lambda=\frac{1}{1+\sqrt{\left(\vec{c}^H\matrix{\Gamma}_{01}\vec{c}\right)^2+\left(\vec{c}^H\matrix{\Gamma}_{02}\vec{c}\right)^2+\left(\vec{c}^H\matrix{\Gamma}_{03}\vec{c}\right)^2}}.
		\label{eq:lambdacond1}
\end{equation}

To satisfy the second transmittance condition we repeat the previous step with \eqref{eq:cond22}, resulting
\begin{equation}
	\lambda=\frac{1}{1+\sqrt{\left(\vec{c}^H\matrix{\Gamma}_{10}\vec{c}\right)^2+\left(\vec{c}^H\matrix{\Gamma}_{20}\vec{c}\right)^2+\left(\vec{c}^H\matrix{\Gamma}_{30}\vec{c}\right)^2}}.
	\label{eq:lambdacond2}
\end{equation}

Examining \eqref{eq:lambdacond1} and \eqref{eq:lambdacond2}, we can stablish the following relationship
\begin{multline}
		\left(\vec{c}^H\matrix{\Gamma}_{01}\vec{c}\right)^2+\left(\vec{c}^H\matrix{\Gamma}_{02}\vec{c}\right)^2+\left(\vec{c}^H\matrix{\Gamma}_{03}\vec{c}\right)^2\\=\left(\vec{c}^H\matrix{\Gamma}_{10}\vec{c}\right)^2+\left(\vec{c}^H\matrix{\Gamma}_{20}\vec{c}\right)^2+\left(\vec{c}^H\matrix{\Gamma}_{30}\vec{c}\right)^2,
\end{multline}
which can be further reduced to
\begin{equation}
	\begin{split}
		\sum_{j=1}^3\left(\vec{c}^H\matrix{\Gamma}_{0j}\vec{c}\right)^2&=\sum_{j=1}^3\left(\vec{c}^H\matrix{\Gamma}_{j0}\vec{c}\right)^2\\
		\sum_{j=1}^3\left(\vec{c}^H\matrix{\Gamma}_{0j}\vec{c}\right)^2&=\sum_{j=1}^3\left(\vec{c}^H\matrix{\Gamma}_{0j}^*\vec{c}\right)^2,
		\label{eq:gammacond}
	\end{split}
\end{equation}
where the last step is applied by the fact that $\matrix{\Gamma}_{nm}=\matrix{\Gamma}_{mn}^*$. One solution is the trivial $\vec{c}=\vec{0}$, but it implies that the Mueller matrix is the zero matrix (no reflection/refraction). Since $\matrix{\Gamma}_{0j}\neq\matrix{\Gamma}_{0j}^*$ and the sum of squares is always greater than $0$, the only possible solution is that $\vec{c}^H\matrix{\Gamma}_{0j}\vec{c}=\vec{c}^H\matrix{\Gamma}_{0j}^*\vec{c}=0,\ j\in[1,2,3]$.

By expanding \eqref{eq:gammadef}, we can observe that
\begin{equation}
	\begin{split}
		\matrix{\Gamma}_{0j}&=\matrix{A}\left(\matrix{\sigma}_0\otimes\matrix{\sigma}_j^*\right)\matrix{A}^{-1}\\
		&=\matrix{A}\begin{pmatrix}\matrix{\sigma}_j^* & \matrix{0}\\\matrix{0} & \matrix{\sigma}_j^*\end{pmatrix}\matrix{A}^{-1},
	\end{split}
\end{equation}
with the following expansions
\begin{equation}
	\begin{split}
		\matrix{\Gamma}_{01}&=\matrix{\Gamma}_{10}^*=2\begin{pmatrix}0 & 1 & 0 & 0\\1 & 0 & 0 & 0\\0 & 0 & 0 & j\\0 & 0 & -j & 0\end{pmatrix}\\
		\matrix{\Gamma}_{02}&=\matrix{\Gamma}_{20}^*=2\begin{pmatrix}0 & 0 & 1 & 0\\0 & 0 & 0 & -j\\1 & 0 & 0 & 0\\0 & j & 0 & 0\end{pmatrix}\\
		\matrix{\Gamma}_{03}&=\matrix{\Gamma}_{30}^*=2\begin{pmatrix}0 & 0 & 0 & 1\\0 & 0 & 1 & 0\\0 & -j & 0 & 0\\j & 0 & 0 & 0\end{pmatrix}.
		\label{eq:gammaexpansions}
	\end{split}
\end{equation}

By defining $\vec{c}=\begin{pmatrix}c_0 & c_1 & c_2 & c_3\end{pmatrix}^T$, we can plug \eqref{eq:gammaexpansions} into \eqref{eq:gammacond} to obtain the following identities:
\begin{equation}
	\begin{split}
		\vec{c}^H\matrix{\Gamma}_{01}\vec{c}&=2\left(c_0^*c_1+c_1^*c_0+j\left(c_2^*c_3-c_3^*c_2\right)\right)\\
		&=4\left(|c_0||c_1|\cos\theta_{0\rightarrow1}+|c_2||c_3|\sin\theta_{2\rightarrow3}\right)\\
		&=0\\
		\vec{c}^H\matrix{\Gamma}_{02}\vec{c}&=2\left(c_0^*c_2+c_2^*c_0+j\left(c_1^*c_3-c_3^*c_1\right)\right)\\
		&=4\left(|c_0||c_2|\cos\theta_{0\rightarrow2}+|c_1||c_3|\sin\theta_{1\rightarrow3}\right)\\
		&=0\\
		\vec{c}^H\matrix{\Gamma}_{03}\vec{c}&=2\left(c_0^*c_3+c_3^*c_0+j\left(c_1^*c_2-c_2^*c_1\right)\right)\\
		&=4\left(|c_0||c_3|\cos\theta_{0\rightarrow3}+|c_1||c_2|\sin\theta_{1\rightarrow2}\right)\\
		&=0,
		\label{eq:c_conds}
	\end{split}
\end{equation}
where $\theta_{i\rightarrow j}=\angle c_i-\angle c_j$. The solution is obtained when $c_0=0$ \underline{and} $\theta_{i\rightarrow j}=0$ ($\angle c_i=\angle c_j),\ i,j\in[1,2,3]$). Hence, the vector $\vec{c}$ shall have the aspect of $\vec{c}=\begin{pmatrix}0&c_1&c_2&c_3\end{pmatrix}^T$, where $\angle c_1=\angle c_2=\angle c_3$. 
\end{proof}

By employing this solution, it becomes evident that $\lambda=1$, thereby yielding $\matrix{C}=\vec{c}\vec{c}^H$. It is noteworthy to highlight that this solution satisfies the eigenvalue condition \eqref{eq:cond1} and the transmittance conditions \eqref{eq:cond21}, \eqref{eq:cond22} ($\lambda>0$, $g_f=g_r=1$).

Despite $\vec{c}\in\mathbb{C}^4$, the aforementioned solution restricts it to possess only $4$ degrees of freedom (rather than $8$). For example, generating $4$ random real numbers ($k_0$, $k_1$, $k_2$, $k_3$) allows for the creation of a coherency matrix and a golden Mueller matrix of the following form:
\begin{equation}
	\begin{split}
	\matrix{C}&=\begin{pmatrix}0 & 0 & 0 & 0\\0 & |c_1|^2 & c_1c_2^* & c_1c_3^*\\0 & c_2c_1^* & |c_2|^2 & c_2c_3^*\\0 & c_3c_1^* & c_3c_2^* & |c_3|^2\end{pmatrix}=\begin{pmatrix}0 & 0 & 0 & 0\\0 & & &\\ 0 & & \matrix{\bar{C}} &\\ 0 & & &\end{pmatrix}\\
	\matrix{M}&=\begin{pmatrix}1 & 0 & 0 & 0\\0 & & &\\ 0 & & \matrix{\bar{M}} &\\ 0 & & &\end{pmatrix}.
	\label{eq:shape_C}
\end{split}
\end{equation}

By adopting this particular design, we can affirm that the generation of only $4$ random numbers enables the synthesis of a golden Mueller matrix that fulfills all the requisite conditions. Furthermore, the structure of $\matrix{M}$ also adheres to \eqref{eq:cond3} and guarantees the invertibility of all golden Mueller matrices.

Consequently, we can define the secret pattern as a vector comprising $4$ random numbers.
\begin{equation}
	\vec{K}=\begin{pmatrix}k_0\\k_1\\k_2\\k_3\end{pmatrix}\in\mathbb{R}^4.
\end{equation}

\begin{algorithm}
	\caption{Secret pattern initialization}
	\begin{algorithmic}[1]
		\Require $\vec{K}=\begin{pmatrix}k_0 & k_1 & k_2 & k_3\end{pmatrix}^T\in\mathbb{R}^4$
		\Ensure $\matrix{M}_{\vec{K}}$
		\State $\vec{c}\leftarrow \begin{pmatrix}0 & k_1 & k_2 & k_3\end{pmatrix}^T\times \exp\left(j \times k_0\right)$.
		\State $\vec{\bar{c}}\leftarrow\vec{c}/\|\vec{c}\|$.
		\State $\matrix{C}\leftarrow\vec{\bar{c}}\vec{\bar{c}}^H$.
		\State $\matrix{M}_{\vec{K}}\leftarrow\left[M_{nm}=\vec{c}^H\matrix{\Gamma}_{nm}\vec{c}\right]$.	
	\end{algorithmic}
\end{algorithm}

Concerning the robustness of Golden Encipherment, we present the following lemma:
\begin{lemma}
	All golden Mueller matrices are strong, secure and achieve the maximum amount of transformation $\mathcal{Q}$.
	\label{lemma:pure_mueller}
\end{lemma}
\begin{proof}
	By applying Lemma \ref{lemma:mueller_trace}, we aim at proving that $\tr\left(\matrix{M}\right)=0$ for all golden Mueller matrices. Thus, by using \eqref{eq:cohe2muepure}, the trace can be described as
	\begin{equation}
		\begin{split}
			\tr\left(\matrix{M}\right)&=M_{00}+M_{11}+M_{22}+M_{33}\\
			&=\vec{c}^H\matrix{\Gamma}_{00}\vec{c}+\vec{c}^H\matrix{\Gamma}_{11}\vec{c}+\vec{c}^H\matrix{\Gamma}_{22}\vec{c}+\vec{c}^H\matrix{\Gamma}_{33}\vec{c}\\
			&=\vec{c}^H\left(\matrix{\Gamma}_{00}+\matrix{\Gamma}_{11}+\matrix{\Gamma}_{22}+\matrix{\Gamma}_{33}\right)\vec{c}.
		\end{split}
	\label{eq:tr_pure_mueller}
	\end{equation}
Matrices $\matrix{\Gamma}_{nn}$ are derived from Pauli matrices, which present interesting properties. In the particular case of $\Gamma_{nn}$ is straightforward to observe that
\begin{equation}
	\matrix{\Gamma}_{00}+\matrix{\Gamma}_{11}+\matrix{\Gamma}_{22}+\matrix{\Gamma}_{33}=\begin{pmatrix}
		4 & 0 & 0 & 0\\0 & 0 & 0 & 0\\0 & 0 & 0 & 0\\0 & 0 & 0 & 0
	\end{pmatrix}.
\end{equation}

Thus, \eqref{eq:tr_pure_mueller} can be further reduced to
\begin{equation}
	\tr\left(\matrix{M}\right)=4\left|c_0\right|^2=0,
\end{equation}
since $\vec{c}=\begin{pmatrix}0 & c_1 & c_2 & c_3\end{pmatrix}^T$.
\end{proof}

In the next section we discuss an alternative to reduce the pattern length.

\subsection{ROTATION ENCIPHERMENT}
\label{sect:rot}
For secure transmission of information between a transmitter and receiver, it is crucial that both parties share and possess knowledge of the alphabet, or constellation, being utilized. An eavesdropper lacking knowledge of the constellation would be unable to accurately demap the received symbols back into the original bits. To further obfuscate the mapping process from potential eavesdroppers, this section employs an arbitrary rotation of the constellation based on secret angles.

Unlike conventional two-dimensional constellations such as PSK or QAM, which only permit rotation about a single axis, Sphere Modulation introduces two additional degrees of freedom. Consequently, the encipherment process entails a rotation of the original reduced Stokes vector into a new vector through the following operation:

\begin{equation}
	\vec{\bar{S}}'=\matrix{R}\vec{\bar{S}},
\end{equation}

where $\matrix{R}$ represents the rotation matrix. The expression for the rotation matrix $\matrix{R}$ can be derived from the Rodrigues rotation formula \cite{Kovacs2012,Rodrigues1840}, which is defined in terms of the rotation angle $\theta$ and the rotation vector $\vec{n}$. The rotation vector $\vec{n}$ is an arbitrary vector perpendicular to the surface of the sphere, thereby defining the orientation of the rotation. The rotation angle $\theta\in\left[0,2\pi\right)$ specifies the number of radians by which the sphere rotates about the rotation vector acting as the rotation axis.

The rotation vector resides on a unit sphere and can be represented in either Cartesian or spherical coordinates.
\begin{equation}
	\begin{split}
		\vec{n}&=\left(n_x,n_y,n_z\right)\\
		&=\left(x,y,\sqrt{1-x^2-y^2}\right)\\
		&=\left(\sin\alpha\cos\beta,\sin\alpha\sin\beta,\cos\alpha\right),
	\end{split}
\label{eq:N_vector}
\end{equation}
where $x,y$ and $\alpha,\beta$ are the coordinates in the Cartesian and spherical systems, respectively. Note that $x^2+y^2=1$, $\alpha\in\left[0,\pi\right)$ and $\beta\in\left[0,2\pi\right)$. Hence, the rotation matrix, by applying the Rodrigues formula \cite{Kovacs2012} is expressed as
\begin{equation}
	\matrix{R}=\matrix{I}+\sin\theta\matrix{N}+\left(1-\cos\theta\right)\matrix{N}^2,
	\label{eq:r_formula}
\end{equation}
where $\matrix{N}$ is the cross-product matrix of the rotation vector $\vec{n}$ and is defined as
\begin{equation}
	\matrix{N}=	\begin{pmatrix}
		0 & -n_z & n_y\\
		n_z & 0 & -n_x\\
		-n_y & n_x & 0
	\end{pmatrix}.
\label{eq:N_matrix}
\end{equation}

By examining \eqref{eq:N_vector} and \eqref{eq:N_matrix}, it becomes evident that the proposed system possesses $3$ degrees of freedom, namely $\alpha$, $\beta$, and $\theta$. Consequently, by defining a secret pattern comprising $3$ random and confidential numbers, we can perform clandestine rotations of the sphere constellation. However, it is imperative to ensure that $\alpha$, $\beta$, and $\theta$ are sufficiently large to prevent minor rotations, which would result in accurate symbol decoding. Generally, the rotation becomes noticeable when the symbols are displaced beyond the noise deviation $\sigma_w$.

In terms of the resilience of the Rotation Encipherment, we establish the following lemma:
\begin{lemma}
	The patterns of the Rotation Encipherment are strong, secure and achieve the maximum amount of transformation for a rotation of $180\deg$.
	\label{lemma:rot_mueller}
\end{lemma}
\begin{proof}
	From Lemma \ref{lemma:mueller_trace}, to prove that $\tr\left(\matrix{M}\right)=0$, we denote the Mueller matrix of Rotation Encipherment as
	\begin{equation}
		\matrix{M}=\begin{pmatrix}
			1 & 0 & 0 & 0\\
			0 & & &\\
			0 & & \matrix{R} & \\
			0 & & &
		\end{pmatrix}.
	\end{equation}
Thus,
	\begin{equation}
	\tr\left(\matrix{M}\right)=1+\tr\left(\matrix{R}\right).
	\label{eq:tr_mueller_rot}
	\end{equation}
By plugging \eqref{eq:r_formula} and \eqref{eq:N_matrix} in \eqref{eq:tr_mueller_rot}, it can be rewritten as
\begin{equation}
	\begin{split}
		\tr\left(\matrix{M}\right)&=1+\tr\left(\matrix{I}\right)+\sin\theta\tr\left(\matrix{N}\right)+\left(1-\cos\theta\right)\tr\left(\matrix{N}^2\right)\\
		&=4+\left(1-\cos\theta\right)\tr\left(\matrix{N}^2\right).
	\end{split}
\label{eq:tr_mueller_rot2}
\end{equation}

The square of the cross-product matrix is equal to
\begin{equation}
	\matrix{N}^2=(-1)\begin{pmatrix}
		n_y^2+n_z^2 & -n_xn_y & -n_xn_z\\
		-n_xn_y & n_x^2+n_z^2 & -n_yn_z\\
		-n_xn_z & -n_yn_z & n_x^2+n_y^2
	\end{pmatrix}.
\end{equation}
Therefore, \eqref{eq:tr_mueller_rot2} can be further reduced to
\begin{equation}
	\begin{split}
		\tr\left(\matrix{M}\right)&=4+\left(1-\cos\theta\right)\tr\left(\matrix{N}^2\right)\\
		&=4-2\left(1-\cos\theta\right)\|\vec{n}\|^2\\
		&=2\left(1+\cos\theta\right)
	\end{split}
\end{equation}
and equals to $0$ when $\theta=\pi$.
\end{proof}

\subsection{OPPOSITE ENCIPHERMENT}
\label{sect:oppo}
In the preceding section, we examined the structure of the Mueller matrix to be a golden Mueller matrix. It was observed that a golden Mueller matrix consists of a $3\times 3$ sub-matrix. The simplest form of a Mueller matrix is the identity matrix; however, in this case, no encryption is performed. Diagonal matrices are another possible solution, but they must satisfy the transmittance and eigenvalue conditions. The transmittance condition is always satisfied when $\matrix{C}$ and $\matrix{M}$ exhibit the form specified in \eqref{eq:shape_C}. This condition implies that $M_{00}^2=1$ and that \eqref{eq:ineqeigen} is satisfied with an equal sign, along with a single positive eigenvalue. Consequently, we can express \eqref{eq:ineqeigen} as follows:
\begin{equation}
	\frac{1}{4}\left(\matrix{M}^T\matrix{M}\right)=\tr\left(\matrix{C}^2\right)=\lambda^2=\tr\left(\matrix{C}\right)^2=M_{00}^2=1.
\end{equation}

By considering a diagonal $\matrix{M}$ with a single eigenvalue, specifically $\lambda=1$, the resulting $\bar{\matrix{M}}$ matrix is also diagonal, with elements of either $1$ or $-1$ along its diagonal. By generating $3$ random bits, we can generate a diagonal golden Mueller matrix using the "Opposite" algorithm. This algorithm multiplies the Stokes parameters by either $1$ (leaving them unchanged) or $-1$ (flipping them to the opposite direction).

In terms of strength analysis, it is evident that the pattern is robust and secure only when $\matrix{\bar{M}}$ contains two $-1$ elements and a single $1$ element along its diagonal, regardless of their positions. It is important to note that the specific case where the diagonal consists entirely of $-1$ elements does not correspond to a pure Mueller matrix, as it fails to satisfy the eigenvalue condition and therefore lacks equivalence in the physical domain.

However, it is worth noting that this design does not always yield a pure Mueller matrix and, consequently, it may exhibit multiple positive/negative eigenvalues. Despite this, the transmittance condition must always be met to preserve energy transmission and polarization degree. As we are considering a diagonal Mueller matrix, the coherency matrix is also diagonal. Utilizing \eqref{eq:gammadef}, we can express the coherency matrix as follows:
\begin{equation}
	\matrix{C}=\frac{1}{4}\left(\matrix{\Gamma}_{00}\pm\matrix{\Gamma}_{11}\pm\matrix{\Gamma}_{22}\pm\matrix{\Gamma}_{33}\right).
\end{equation}

Furthermore, as it is diagonal, the eigenvalues are the same as the diagonal values. This problem has three solutions:
\begin{subequations}
	\begin{align}
		M_{00}=1, M_{11}=1, M_{22}=-1, M_{33}=-1\\
		M_{00}=1, M_{11}=-1, M_{22}=1, M_{33}=-1\\
		M_{00}=1, M_{11}=-1, M_{22}=-1, M_{33}=1,
	\end{align}
\end{subequations}
which all produce a diagonal coherence matrix.

\section{IMPERFECT POLARIZATION ANALYTICAL ANALYSIS}
\label{sect:imp_pol}
In the preceding sections, we examined the security of the physical layer using polarization techniques. All the schemes presented thus far assumed perfect calibration between orthogonal polarizations in order to establish the mapping between Jones and Stokes vectors, as well as the Poincaré sphere. However, in practical implementations, perfect polarization is an idealized concept that is not easily achievable. In this section, we will analyze the effects of imperfect polarizations.

Imperfect polarizations can be modeled using a matrix that modifies the incident Jones vector as follows:
\begin{equation}
	\vec{\tilde{E}}_0=\matrix{Q}\vec{E}_0,
\end{equation}
where $\matrix{Q}$ is defined as
\begin{equation}
	\matrix{Q}=\begin{pmatrix}
		1 & a\\b & c
	\end{pmatrix},
\end{equation}
where $a,b,c\in\mathbb{C},\ |a|^2,|b|^2,|c|^2\leq 1$. Depending on $a,b,c$ factors, we are able to particularize the imperfect polarization to cross-polarization or unbalanced polarizations, for instance. 

The $\matrix{Q}$ matrix represents the imperfect polarization and is a Jones matrix that alters the incident polarization state. It can also be associated with a Mueller matrix denoted as $\matrix{M}_{\matrix{Q}}$, which is computed using \eqref{eq:mueller_orig}. The expression for the imperfect Mueller matrix is given by \eqref{eq:gen_imperf_pol}. This imperfect Mueller matrix linearly modifies the incident Stokes vector, resulting in the transformed Stokes vector given by
\begin{equation}
	\vec{\tilde{S}}=\matrix{M}_{\matrix{Q}}\vec{S}'=\matrix{M}_{\matrix{Q}}\matrix{M}_{\vec{K}}\vec{S}.
\end{equation}

\begin{table*}
	\centering
	\begin{minipage}{0.85\textwidth}
		\begin{align}
	\matrix{M}_{\matrix{Q}}=\begin{pmatrix}
		\frac{1}{2}\left(1+|a|^2+|b|^2+|c|^2\right) & \frac{1}{2}\left(1+|b|^2-|a|^2-|c|^2\right) & \Re\left\{a+bc^*\right\} & -\Im\left\{a-bc^*\right\}\\
		\frac{1}{2}\left(1+|a|^2-|b|^2-|c|^2\right) & \frac{1}{2}\left(1+|c|^2-|a|^2-|b|^2\right) & \Re\left\{a-bc^*\right\} & -\Im\left\{a+bc^*\right\}\\
		\Re\left\{b+ac^*\right\} & \Re\left\{b-ac^*\right\} & \Re\left\{c+ab^*\right\} & -\Im\left\{c+ab^*\right\}\\
		\Im\left\{b-ac^*\right\} & \Im\left\{b+ac^*\right\} & \Im\left\{c-ab^*\right\} & \Re\left\{c-ab^*\right\}
	\end{pmatrix}
\label{eq:gen_imperf_pol}
\end{align}
\hrule
\end{minipage}
\end{table*}

Depending on the values of $a,b,c$ we particularize the following imperfections.

\subsection{CROSS-POLARIZATION}
One of the commonly encountered phenomena is the cross-polarization effect. In practice, RF chains are not perfectly isolated, and antennas may not be perfectly orthogonal, leading to coupling between orthogonal polarizations. This can be modeled by setting $a=b=\xi$ and $c=1$, resulting in the imperfect Mueller matrix expression of:
\begin{equation}
	\matrix{Q}=\begin{pmatrix}
		1 & \xi\\\xi & 1
	\end{pmatrix},
\label{eq:Q_xpol}
\end{equation}
where $\xi\in\mathbb{C},\ |\xi|^2\leq 1$ is the cross-polarization factor.

Particularizing \eqref{eq:gen_imperf_pol} with $a=b=\xi$ and $c=1$, we obtain
\begin{equation}
	\begin{split}
	\matrix{M}_{\matrix{Q}}&=\begin{pmatrix}
		1+|\xi|^2 & 0 & 2\Re\left\{\xi\right\} & 0\\
		0 & 1-|\xi|^2 & 0 & -2\Im\left\{\xi\right\} \\
		2\Re\left\{\xi\right\} & 0 & 1+|\xi|^2 & 0\\
		0 & 2\Im\left\{\xi\right\} & 0 & 1-|\xi|^2
	\end{pmatrix} 
\\
\tilde{S}_0 & = \left(1+|\xi|^2\right)S_0+2\Re\left\{\xi\right\}S_2\\
\tilde{S}_1 & = \left(1-|\xi|^2\right)S_1-2\Im\left\{\xi\right\}S_3\\
\tilde{S}_2 & = \left(1+|\xi|^2\right)S_2+2\Re\left\{\xi\right\}S_0\\
\tilde{S}_3 & = \left(1-|\xi|^2\right)S_3+2\Im\left\{\xi\right\}S_1\\
\end{split}
\end{equation}

It is interesting to observe that when $\xi\rightarrow 1$, the resulting Stokes parameters collapse to
\begin{equation}
	\begin{split}
		\tilde{S}_0&=2S_0+2S_2\\
		\tilde{S}_1&=0\\
		\tilde{S}_2&=2S_2+2S_0\\
		\tilde{S}_3&=0.
	\end{split}
\end{equation}

This phenomenon results in the combination of both polarizations, resulting in an oblique polarization (with a slant of 45 degrees). In contrast, when $\xi\rightarrow j$, the Stokes parameters collapse to:

\begin{equation}
	\begin{split}
		\tilde{S}_0&=2S_0\\
		\tilde{S}_1&=-2S_3\\
		\tilde{S}_2&=2S_2\\
		\tilde{S}_3&=2S_1.
	\end{split}
\end{equation}
In this case, the $S_2$ parameter (representing the slant) remains unchanged, while $S_1$ and $S_3$ undergo a flip. The $S_1$ parameter measures the verticality/horizontality of polarization, and the $S_3$ parameter measures the eccentricity of the polarization ellipse. The exchange between $S_1$ and $S_3$ implies that linear polarizations are transformed into circular polarizations, and vice versa.

\subsection{UNBALANCED POLARIZATION}
In this case, one polarization is more attenuated compared with the other. The imperfect matrix is denoted by
\begin{equation}
	\matrix{Q}=\begin{pmatrix}
		1 & 0\\0 & \xi
	\end{pmatrix},
\label{eq:Q_unbal}
\end{equation}
and thus,
\begin{equation}
	\begin{split}
		\matrix{M}_{\matrix{Q}}&=\frac{1}{2}\begin{pmatrix}
			1+|\xi|^2 & 1-|\xi|^2 & 0 & 0\\
		    1-|\xi|^2 & 1+|\xi|^2 & 0 & 0 \\
			0 & 0 & 2\Re\left\{\xi\right\} & -2\Im\left\{\xi\right\} \\
			0 & 0 & 2\Im\left\{\xi\right\} & 2\Re\left\{\xi\right\}
		\end{pmatrix} 
		\\
		\tilde{S_0} & = \frac{S_0+S_1}{2}+\frac{S_0-S_1}{2}|\xi|^2\\
		\tilde{S_1} & = \frac{S_0+S_1}{2}+\frac{S_1-S_0}{2}|\xi|^2\\
		\tilde{S_2} & = \Re\left\{\xi\right\}S_2-\Im\left\{\xi\right\}S_3\\
		\tilde{S_3} & = \Im\left\{\xi\right\}S_2+\Re\left\{\xi\right\}S_3\\
	\end{split}
\end{equation}
When $\xi\rightarrow 1$, the imperfect matrix approaches the identity matrix, indicating no polarization imbalance. However, when $\xi\rightarrow j$, a phase shift is introduced to the secondary polarization. While $S_0$ and $S_1$ remain unchanged, $S_2$ and $S_3$ are exchanged. As a result, circular polarizations are transformed into slant polarizations, and vice versa.

\subsection{IMPACT OF IMPERFECT POLARIZATION}
The impact of imperfect polarization can be studied using Mueller calculus, which describes how the Mueller matrix used for encryption is modified. Assuming that the imperfections occur at the RF chain, the global Mueller matrix can be represented as:

\begin{equation}
	\matrix{M}_G=\matrix{M}_{\matrix{Q}}\matrix{M}_{\vec{K}}.
\end{equation}

In the case of cross-polarization, and assuming the use of golden encryption matrices, the global Mueller matrix can be expressed as shown in \eqref{eq:glob_xpol}. By assuming that $M_{01}=M_{02}=M_{03}=0$, we observe that both the magnitude $|\xi|^2$ and the imaginary part $\Im\left\{\xi\right\}$ have a significant impact on the global Mueller matrix.
\begin{table*}
	\centering
	\begin{minipage}{0.85\textwidth}
		\begin{align}
	\matrix{M}_G=\begin{pmatrix}
		1+|\xi|^2 & 2M_{21}\Re\left\{\xi\right\} & 2M_{22}\Re\left\{\xi\right\} & 2M_{23}\Re\left\{\xi\right\}\\
		0 & \left(1-|\xi|^2\right)M_{11}-2M_{31}\Im\left\{\xi\right\} & \left(1-|\xi|^2\right)M_{12}-2M_{32}\Im\left\{\xi\right\} & \left(1-|\xi|^2\right)M_{13}-2M_{33}\Im\left\{\xi\right\}\\
		0 & M_{21}\left(1+|\xi|^2\right) & M_{22}\left(1+|\xi|^2\right) & M_{23}\left(1+|\xi|^2\right)\\
		0 & \left(1-|\xi|^2\right)M_{31}+2M_{11}\Im\left\{\xi\right\} & \left(1-|\xi|^2\right)M_{32}+2M_{12}\Im\left\{\xi\right\} & \left(1-|\xi|^2\right)M_{33}+2M_{13}\Im\left\{\xi\right\}
	\end{pmatrix}
\label{eq:glob_xpol}
\end{align}
\hrule
\end{minipage}
\end{table*}

In the case of unbalanced polarizations, the global Mueller matrix can be expressed as shown in \eqref{eq:glob_unbal}. Unlike the case of cross-polarization, when polarizations are unbalanced, the Mueller coefficients are affected only when $\xi$ has an imaginary component. In this scenario, the imaginary part of $\xi$ primarily influences the bottom-right block matrix of the global Mueller matrix, while the top-right block matrix is primarily affected by the magnitude $|\xi|$.
\begin{table*}
	\centering
	\begin{minipage}{0.85\textwidth}
		\begin{align}
			\matrix{M}_G=\frac{1}{2}\begin{pmatrix}
				1+|\xi|^2 & \left(1-|\xi|^2\right)M_{11} & \left(1-|\xi|^2\right)M_{12} & \left(1-|\xi|^2\right)M_{13}\\
				0 & \left(1+|\xi|^2\right)M_{11} & \left(1+|\xi|^2\right)M_{12} & \left(1+|\xi|^2\right)M_{13}\\
				0 & 2M_{21}\Re\left\{\xi\right\}-2M_{31}\Im\left\{\xi\right\} & 2M_{22}\Re\left\{\xi\right\}-2M_{32}\Im\left\{\xi\right\} & 2M_{23}\Re\left\{\xi\right\}-2M_{33}\Im\left\{\xi\right\}\\
				0 & 2M_{31}\Re\left\{\xi\right\}+2M_{21}\Im\left\{\xi\right\} & 2M_{32}\Re\left\{\xi\right\}+2M_{22}\Im\left\{\xi\right\} & 2M_{33}\Re\left\{\xi\right\}+2M_{23}\Im\left\{\xi\right\}\\
			\end{pmatrix}
			\label{eq:glob_unbal}
		\end{align}
		\hrule
	\end{minipage}
\end{table*}

It is of particular interest to note that when $\xi\in\mathbb{R}$, indicating that the unbalance only affects the magnitude of the signal, the global Mueller matrix remains unchanged but scaled. 

It should be emphasized that while the receiver possesses knowledge of $\matrix{M}_{\vec{K}}$ due to the shared obfuscation pattern, the information regarding the imperfect polarization at both the transmitter and receiver ($\matrix{M}_{\matrix{Q}}$) is unknown. Under ideal conditions, where $M_{G10}$, $M_{G20}$, and $M_{G30}$ are all zero, an estimator can be devised if $\xi$ is known to be real. The estimator can be formulated as follows:
\begin{equation}
	\hat{\xi}=\sqrt{1-\frac{1}{3}\left(\frac{M_{G10}}{M_{11}}+\frac{M_{G20}}{M_{12}}+\frac{M_{G30}}{M_{13}}\right)}.
\end{equation}

\section{RESULTS}
In this section we analyze, via simulations, the results of the previous sections under particular circumstances. 

\subsection{SIMULATION METHODOLOGY}
In order to assess the performance of the proposed encryption schemes, we conducted simulations using a carefully designed methodology. The simulation setup consisted of a transmitter and a receiver, both employing spherical modulation and equipped with two orthogonal dipoles. The receiver implemented the Maximum Likelihood (ML) decoding strategy, which is known to be the optimal receiver for this type of modulation.

The channel model employed was Additive White Gaussian Noise (AWGN), where the channel noise was generated by creating $N=10^6$ independent realizations of Gaussian noise with zero mean. The Signal-to-Noise Ratio (SNR) level was adjusted to evaluate the performance under different noise conditions. In some cases, an SNR scanning approach was utilized to analyze the system's performance across a range of SNR values.

For each channel realization, the transmitter generated a new random secret pattern based on one of the three encryption schemes: Golden, Rotation, or Opposite. The secret pattern was then securely exchanged with the legitimate receiver. At the transmitter, the encryption process was performed according to Algorithm \ref{alg:enc}, while the decryption process at the receiver followed Algorithm \ref{alg:dec} to recover the original transmitted data.

During each channel realization, a block of $N_b=64$ bits was transmitted and modulated using a constellation with a default dimension of $M=8$. The ML decoder operated on the spherical constellation and employed a decision rule to determine the closest point on the Poincaré sphere corresponding to the received symbols.

\subsection{AMOUNT OF TRANSFORMATION}
Theorems \ref{theorem:uplowbounds} and Corollaries \ref{theorem:maxdist} and \ref{theorem:finupp} provide valuable insights into the finite magnitude Amount of Transformation, denoted as $\mathcal{Q}$. Additionally, Lemmas \ref{lemma:pure_mueller} and \ref{lemma:rot_mueller} establish that the Amount of Transformation is maximized when the Mueller matrices are golden and when a rotation angle of $\pi$ is applied, respectively.

\begin{figure}[!ht]
\minipage{0.49\textwidth}
	\centering
	\includegraphics[width=0.90\linewidth,clip=true]{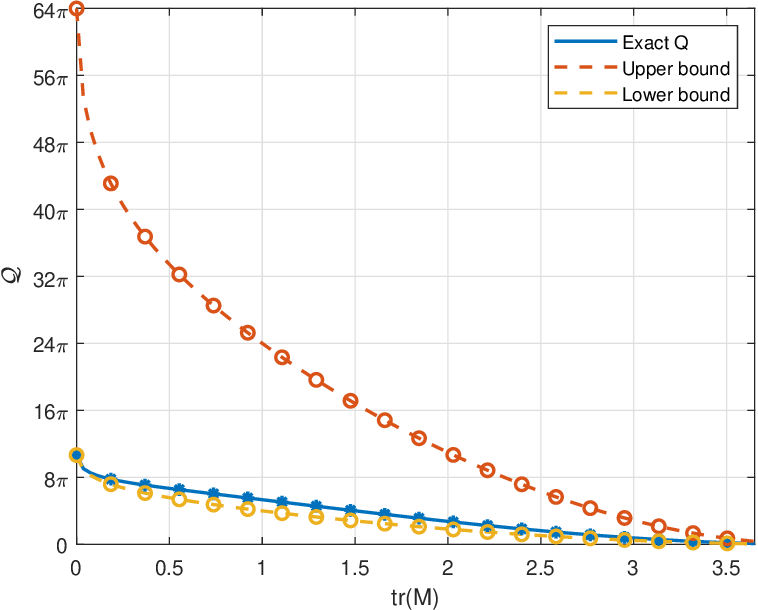}
	\caption{Amount of Transformation $\mathcal{Q}$ as a function of $\tr\left(\underline{\underline{\boldsymbol{M}}}\right)$ for Golden Encipherment.}
	\label{fig:Q_as_trace}
\endminipage\hfill
\minipage{0.49\textwidth}
	\centering
	\includegraphics[width=0.90\linewidth,clip=true]{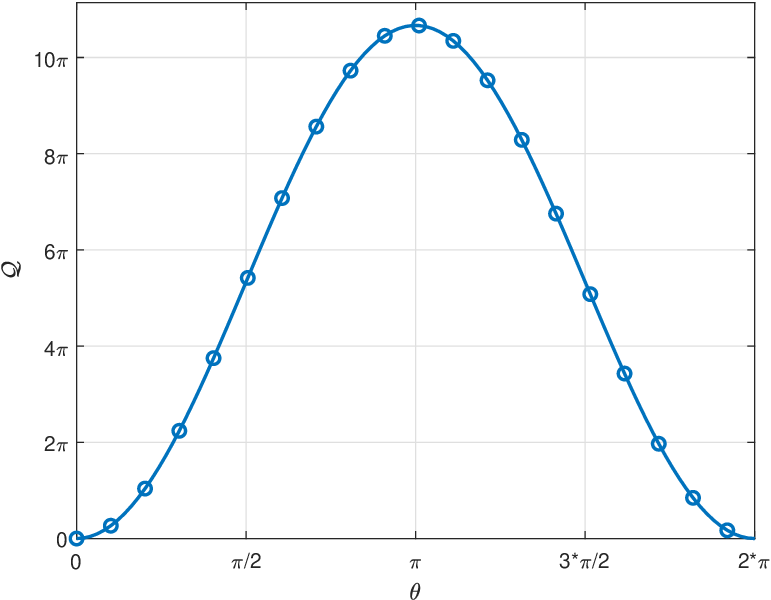}
	\caption{Amount of Transformation $\mathcal{Q}$ as a function of rotation angle $\theta$ for Rotation Encipherment.}
	\label{fig:Q_as_theta}
	\endminipage
\end{figure}

To validate these theoretical results numerically, we present Fig. \ref{fig:Q_as_trace} and Fig. \ref{fig:Q_as_theta}. Fig. \ref{fig:Q_as_trace} illustrates the relationship between the Amount of Transformation $\mathcal{Q}$ and the trace of the Mueller matrix, denoted as $\tr(\matrix{M})$. To generate this figure, we randomly generated $N=10^6$ Mueller matrices without any constraints on the trace (i.e., not necessarily golden). Statistically, as the trace tends to zero, the Amount of Transformation $\mathcal{Q}$ approaches its maximum value.

Furthermore, for each randomly generated Mueller matrix, we computed the corresponding lower bound, $\frac{4\pi}{3}\left\|\matrix{M}-\matrix{I}\right\|_F^2$, and upper bound, $8\pi\left\|\matrix{M}-\matrix{I}\right\|_F^2$, on the Amount of Transformation. As depicted in Fig. \ref{fig:Q_as_trace}, the computed Amount of Transformation $\mathcal{Q}$ falls within the computed lower and upper bounds. Additionally, we can observe the results of Corollary \ref{theorem:finupp}, which states that the maximum achievable Amount of Transformation is $64\pi$.

For the Rotation Encipherment scheme, Fig. \ref{fig:Q_as_theta} showcases the Amount of Transformation $\mathcal{Q}$ as a function of the rotation angle $\theta$. This figure was obtained by generating $N=10^6$ random normal vectors on the surface of the sphere and computing the corresponding Rotation Mueller matrix for various values of the rotation angle $\theta$ using the formula in \eqref{eq:r_formula}. As demonstrated by Lemma \ref{lemma:rot_mueller}, the Amount of Transformation is maximized at $\theta=\pi$ and minimized at $\theta=0$ or $\theta=2\pi$. In the latter cases, the Mueller matrix remains unchanged ($\matrix{M}=\matrix{I}$), resulting in no transformation.

\subsection{ENCIPHERMENT COMPARISON}
Sections \ref{sect:golden}, \ref{sect:rot}, and \ref{sect:oppo} introduced three distinct encipherments: Golden, Rotation, and Opposite. In this section, we compare the performance of these encipherments against an eavesdropper in the presence of an AWGN channel. To conduct a comprehensive analysis, we conducted extensive simulations comprising $N=10^7$ realizations of the AWGN channel for each encipherment scheme. The employed modulation technique is Sphere Modulation, with the constellation points defined in \cite{Henarejos2018} for $M=4$, $M=8$, $M=16$, and $M=32$.

For the Golden Encipherment, we utilized Algorithms \ref{alg:enc} and \ref{alg:dec}, where the legitimate receiver possesses the secret pattern and can perform the inverse operation of the equivalent Mueller matrix on the received Stokes vector. In contrast, the eavesdropper lacks knowledge of the secret pattern and does not perform any Mueller inverse.

The Rotation Encipherment involves rotating the spherical constellation by generating a random normal vector on the surface. To maximize the Amount of Transformation, as expressed in \eqref{eq:def_amount_transformation}, we employed a rotation angle of $\theta=\pi$ radians. With this encryption scheme, we randomly generated a normal vector on the spherical constellation and rotated it by $180\deg$.

Lastly, the Opposite Encipherment is a special case of the Golden Encipherment, where the Mueller matrix is diagonal with its components being $\pm 1$. In Algorithm \ref{alg:dec}, there is no matrix inversion involved; instead, the received Stokes vector is multiplied by the same secret pattern used by the transmitter, which only contains $\pm 1$ elements.

\begin{figure}[!ht]
	\centering
	\subfloat[][$M=4$]{\includegraphics[width=0.45\linewidth,clip=true]{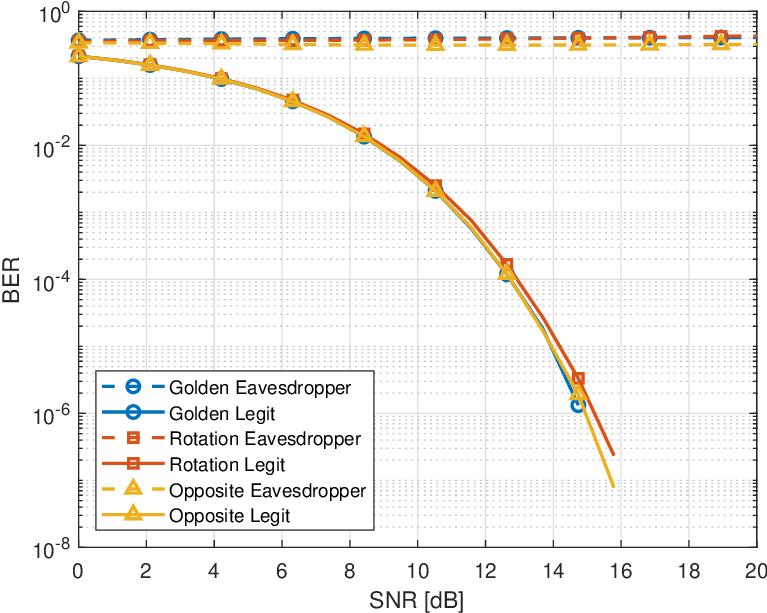}\label{fig:snr_random_mueller_4}}\hfill
	\subfloat[][$M=8$]{\includegraphics[width=0.45\linewidth,clip=true]{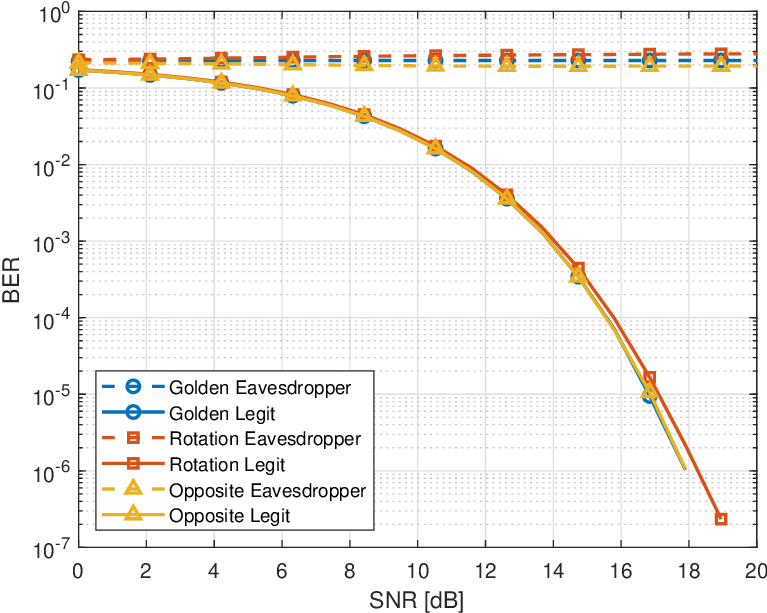}\label{fig:snr_random_mueller_8}}\hfill
	\subfloat[][$M=16$]{\includegraphics[width=0.45\linewidth,clip=true]{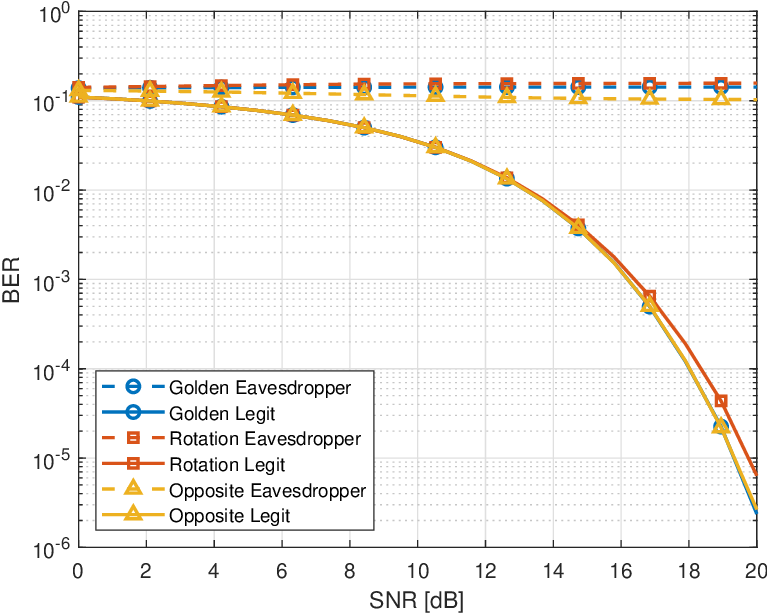}\label{fig:snr_random_mueller_16}}\hfill
	\subfloat[][$M=32$]{\includegraphics[width=0.45\linewidth,clip=true]{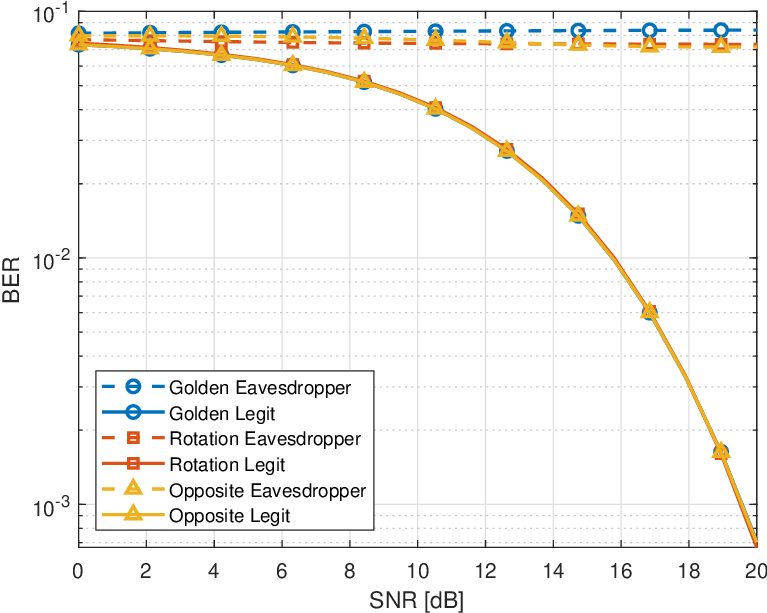}\label{fig:snr_random_mueller_32}}\hfill
	\caption{SNR comparison of Golden, Rotation and Opposite Encipherments in front of an eavesdropper for constellations $M=4$, $M=8$, $M=16$ and $M=32$.}
	\label{fig:snr_random_mueller}
\end{figure}

Fig. \ref{fig:snr_random_mueller} presents the SNR and the corresponding Bit Error Rate (BER) for the Golden, Rotation, and Opposite Encipherments in the presence of an eavesdropper who is unaware of the exchanged secret pattern between the legitimate transmitter and receiver. As expected, the eavesdropper is unable to successfully decode the transmitted information, resulting in the same BER at any SNR. Conversely, the Golden, Rotation, and Opposite Encipherments exhibit similar BER curves, indicating the effectiveness of the encryption/decryption algorithms in successfully decoding the information with knowledge of the secret pattern.

In addition, Fig. \ref{fig:snr_random_mueller} illustrates that all three encipherment schemes achieve the same BER. However, the Opposite Encipherment exhibits a slightly lower BER for an eavesdropper, indicating that it is less robust compared to the other encipherments. Nevertheless, all three encipherments achieve a flat BER curve, independent of the SNR, which is desirable. This characteristic holds true across all modulation orders. Moreover, as the modulation order increases, the BER for an eavesdropper decreases.

\subsection{ROTATION ANGLE ANALYSIS}
\begin{figure}[!ht]
	\centering
	\subfloat[][$M=4$]{\includegraphics[width=0.45\linewidth,clip=true]{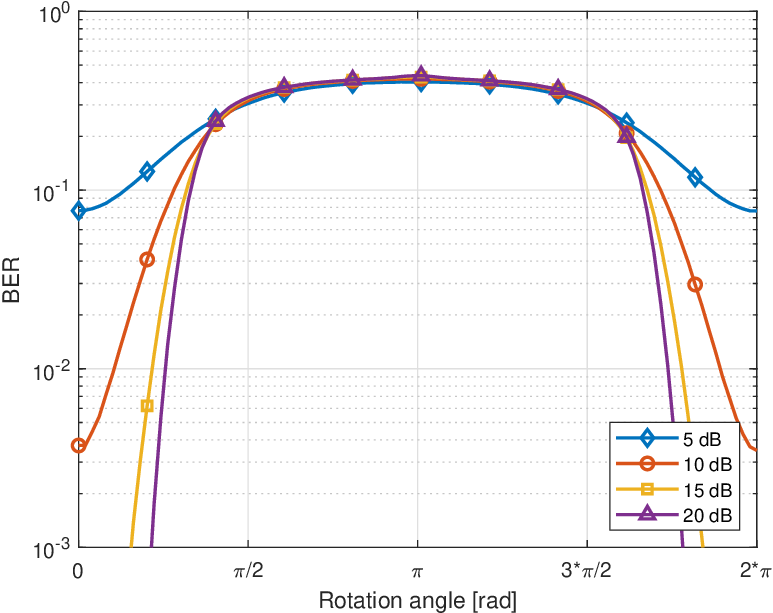}\label{fig:rotation_ber_4}}\hfill
	\subfloat[][$M=8$]{\includegraphics[width=0.45\linewidth,clip=true]{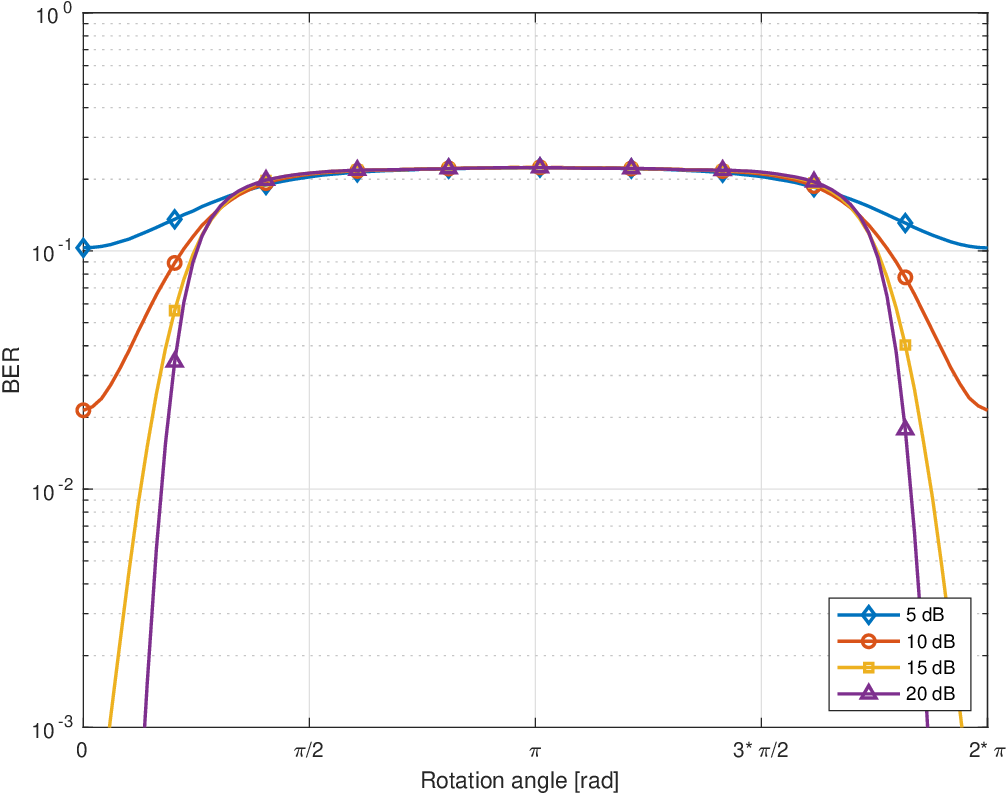}\label{fig:rotation_ber_8}}\hfill
	\subfloat[][$M=16$]{\includegraphics[width=0.45\linewidth,clip=true]{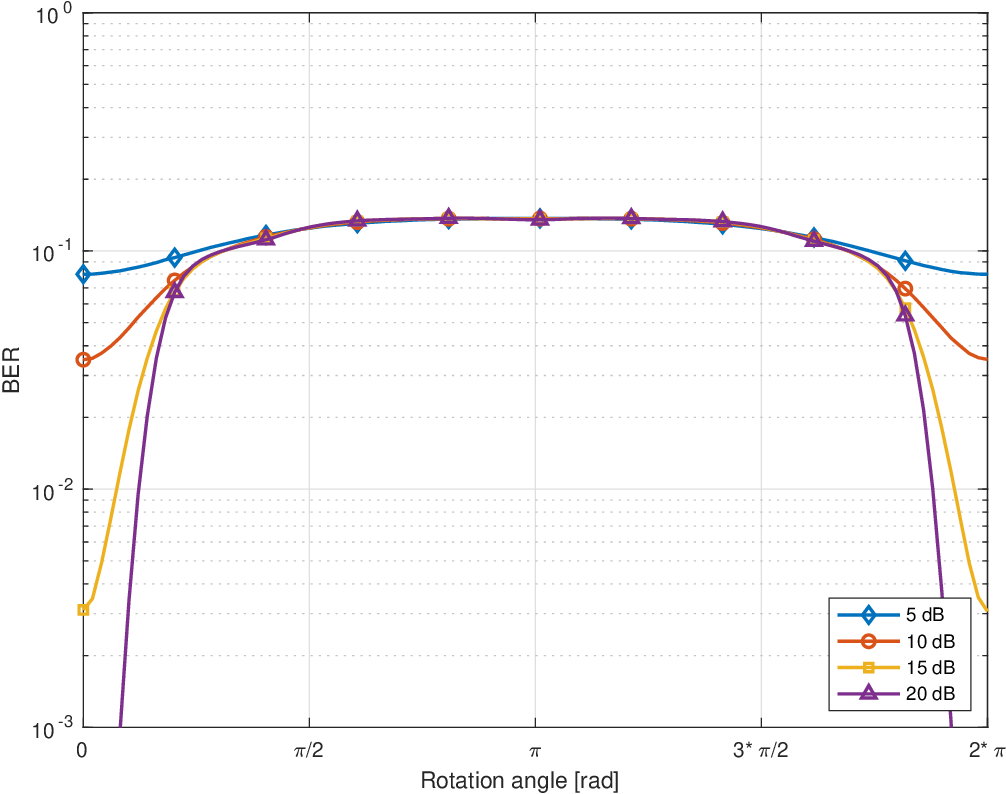}\label{fig:rotation_ber_16}}\hfill
	\subfloat[][$M=32$]{\includegraphics[width=0.45\linewidth,clip=true]{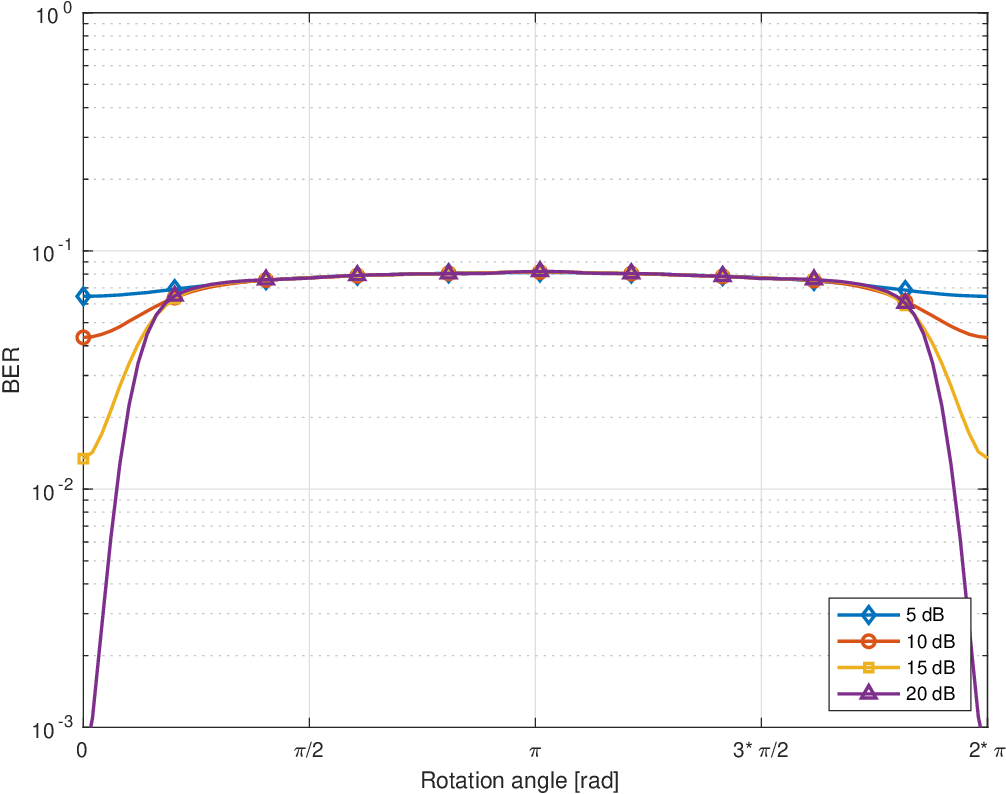}\label{fig:rotation_ber_32}}\hfill
	\caption{BER of an eavesdropper for different rotation angles at SNR=$\SI{5}{dB}$, SNR=$\SI{10}{dB}$, SNR=$\SI{15}{dB}$ and SNR=$\SI{20}{dB}$.}
	\label{fig:rotation_ber}
\end{figure}

The Rotation Encipherment involves rotating the constellation using a random normal vector and a random rotation angle. However, as demonstrated in Lemma \ref{lemma:rot_mueller}, the rotation angle plays a crucial role in achieving the maximum Amount of Transformation. In order to assess the impact of the rotation angle, Fig. \ref{fig:rotation_ber} examines the influence of different rotation angles on the Bit Error Rate (BER) for an eavesdropper across various SNR values. The maximum BER is observed when the Amount of Transformation is at its peak, which occurs at $\theta=\pi$ as shown in Lemma \ref{lemma:rot_mueller}. At this point, the information is transformed to its maximum extent, rendering the eavesdropper incapable of decoding the transmitted data.

Interestingly, Fig. \ref{fig:rotation_ber} reveals a plateau in the BER curve between $\theta=\pi/2$ and $\theta=3\pi/2$, where the eavesdropper's BER remains nearly constant regardless of the rotation angle. This characteristic allows for the utilization of the rotation angle as a third random parameter for pattern generation, as it yields a comparable BER. For instance, the transmitter and receiver could generate a secret pattern based on random angles $\alpha$ and $\beta$ for the normal vector, along with $\theta\in\left[\frac{\pi}{2},\frac{3\pi}{2}\right]$.

This plateau phenomenon becomes more pronounced as the constellation size increases. The reason behind this is that with a larger number of constellation points, the minimum distance between points decreases. As a result, small rotations introduce a higher probability of error. While this effect does not impact the legitimate receiver, as the transformation is reversed, it poses a significant challenge for an eavesdropper, akin to introducing substantial noise that displaces all the constellation points.

\subsection{VARIANCE AND STOKES SNR}
In Sections \ref{sect:stokes_stats} and \ref{sect:stokes_snr}, we provide analytical descriptions of the transformation of signal statistics during the Jones-Stokes conversion. We emphasize that this transformation is nonlinear and affects the different Stokes parameters in distinct ways.

\begin{figure}[!ht]
	\minipage{0.49\textwidth}
	\centering
	\includegraphics[width=0.85\linewidth,clip=true]{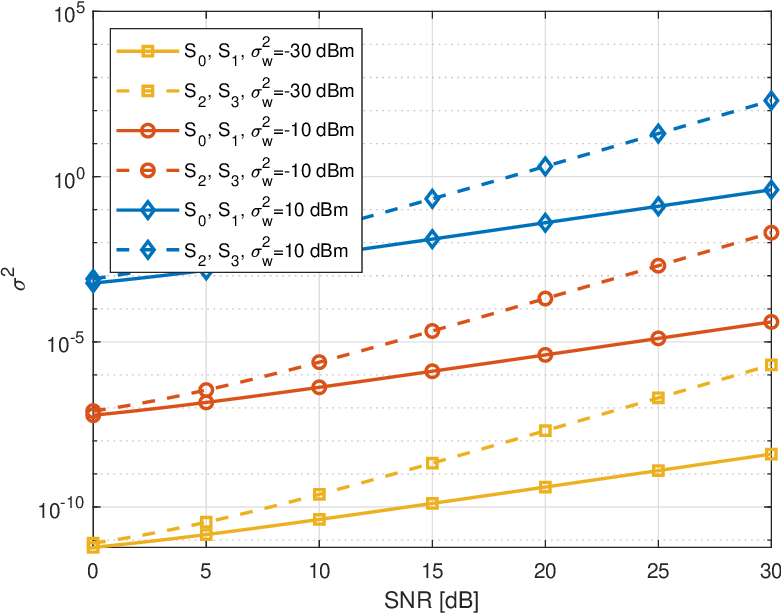}
	\caption{Variance of Stokes Vector depending on the input SNR for $\sigma_w^2=\SI{-30}{dBm}$, $\sigma_w^2=\SI{-10}{dBm}$ and $\sigma_w^2=\SI{10}{dBm}$.}
	\label{fig:var_noise}
	\endminipage\hfill
	\minipage{0.49\textwidth}
	\centering
	\includegraphics[width=0.85\linewidth,clip=true]{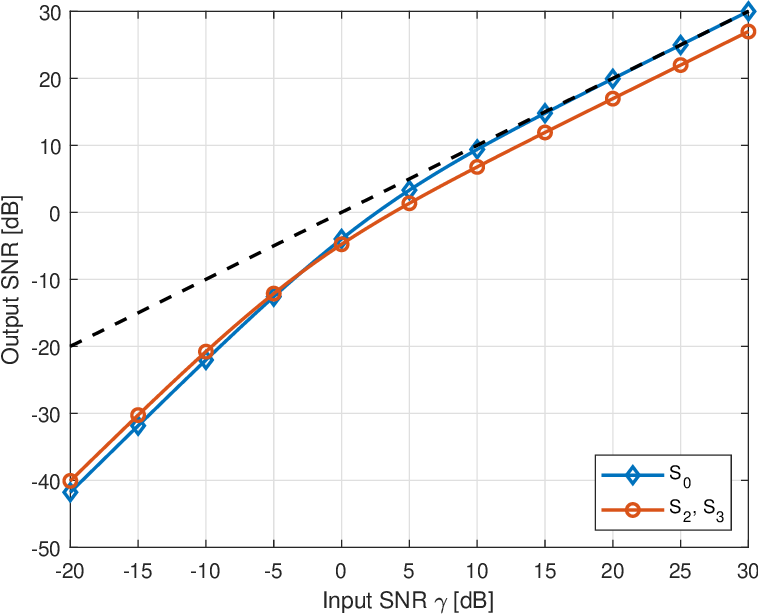}
	\caption{Output SNR after the Jones-Stokes transformation. The dashed line indicates $\textrm{Input SNR}=\textrm{Output SNR}$.}
	\label{fig:snr_stokes}
	\endminipage
\end{figure}

For instance, the variances of $S_2$ and $S_3$ are equal to $S_0$ and $S_1$ plus $2P_x^2$. As a result, the difference between $S_0$ and $S_2$ increases proportionally with the SNR. Fig. \ref{fig:var_noise} presents the variances of the Stokes parameters as a function of input SNR for different levels of noise ($\sigma_w^2=\SI{-30}{dBm}$, $\sigma_w^2=\SI{-10}{dBm}$, and $\sigma_w^2=\SI{10}{dBm}$). Firstly, we observe the increasing difference between $S_0$ and $S_1$ compared to $S_2$ and $S_3$, which grows proportionally with the SNR. Secondly, the variances are proportional to the SNR, driven by the mixed factor $4P_x\sigma_w^2$. Lastly, we note that the variances decrease as the noise level diminishes, as expected.

Regarding the Output SNR after the Jones-Stokes conversion, Fig. \ref{fig:snr_stokes} illustrates its transformation. As described in \eqref{eq:snr_gamma}, the SNR of the Stokes parameters is lower than the input SNR. At low SNR values, the Output SNR is quadratically proportional to $\gamma^2$, resulting in a slope of $2$ in decibels. At high SNR values, it is directly proportional to $\gamma$, yielding a slope of $1$ in decibels. Thus, we conclude that the Output SNR is quadratically proportional in the low SNR regime and directly proportional in the high SNR regime.

Furthermore, in the low SNR regime, the SNR of $S_0$ is lower compared to that of $S_2$ or $S_3$. However, in the high SNR regime, this behavior is reversed, and $\text{SNR}_0\geq\textrm{SNR}_2$. This reversal occurs at $\gamma=\frac{1}{2}$ (or $\gamma=\SI{-3}{dB}$). Consequently, the degradation of SNR is more significant at low SNR regimes.

\subsection{IMPERFECT POLARIZATION}
In Section \ref{sect:imp_pol}, we investigate the impact of imperfect polarization caused by misalignment between transmitting and receiving antennas, which generates cross-polarization, or due to an imbalance between polarizations. Depending on the specific case under study, the imperfect matrix $\matrix{Q}$ takes the form of \eqref{eq:Q_xpol} or \eqref{eq:Q_unbal}.

\begin{figure}[!ht]
	\centering
	\subfloat[][$M=8$]{\includegraphics[width=0.45\linewidth,clip=true]{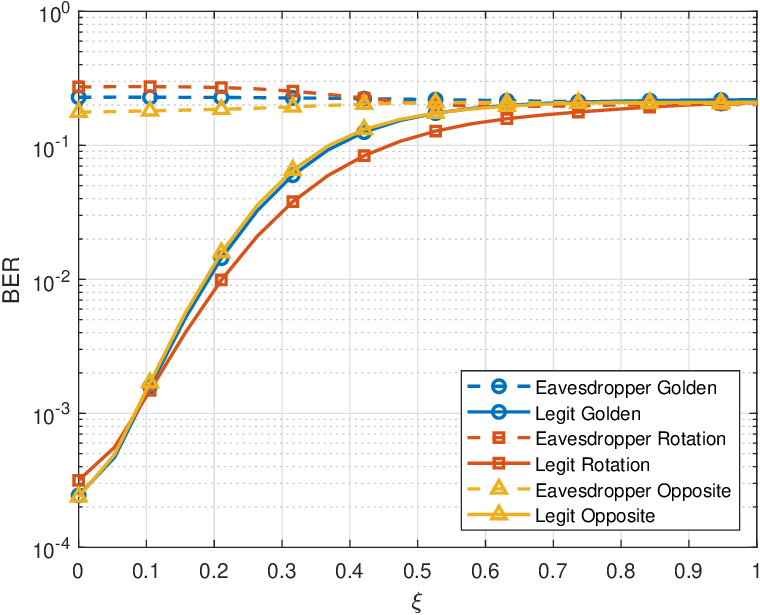}\label{fig:xpol_all_xpol_8_15}}\hfill
	\subfloat[][$M=16$]{\includegraphics[width=0.45\linewidth,clip=true]{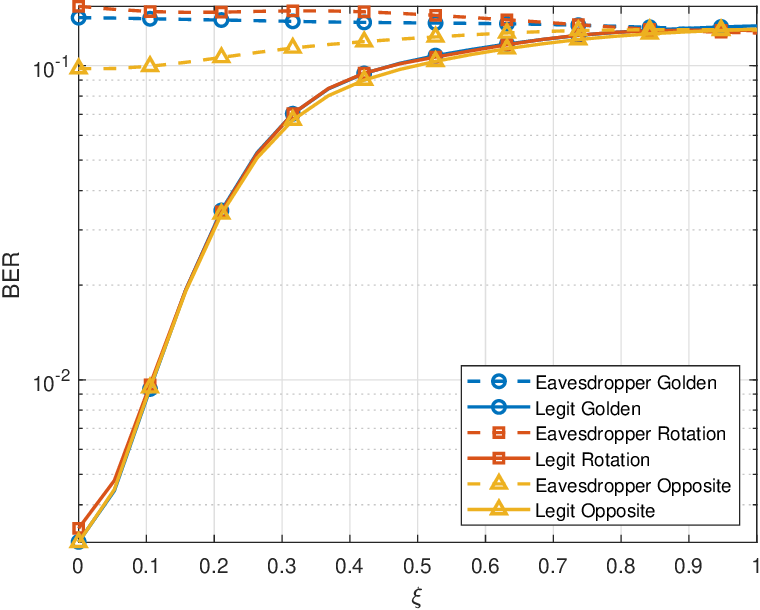}\label{fig:xpol_all_xpol_16_15}}\hfill
	\caption{SNR degradation due to cross polarization as a function of $\xi$ for constellations $M=8$ and $M=16$.}
	\label{fig:xpol_all_xpol}
\end{figure}

Fig. \ref{fig:xpol_all_xpol} illustrates the degradation of SNR as a function of $\xi$ in the case of cross-polarization. As $\xi\rightarrow 1$, the degradation is maximized. At this level, cross-polarization is at its highest, and there is no distinction between the two polarizations. This situation occurs, for example, when the antennas are rotated by $\SI{45}{\deg}$ at either the transmitter or receiver side.

\begin{figure}[!ht]
	\centering
	\subfloat[][$M=8$]{\includegraphics[width=0.45\linewidth,clip=true]{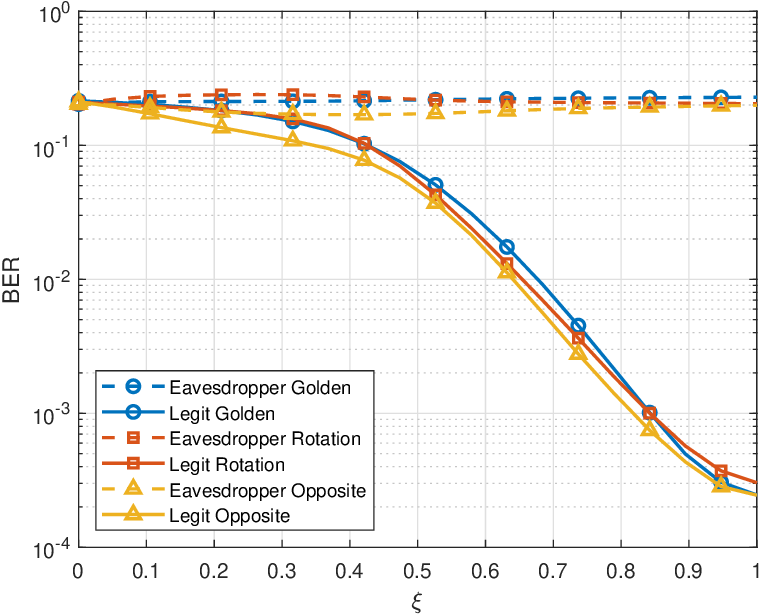}\label{fig:xpol_all_unbalance_8_15}}\hfill
	\subfloat[][$M=16$]{\includegraphics[width=0.45\linewidth,clip=true]{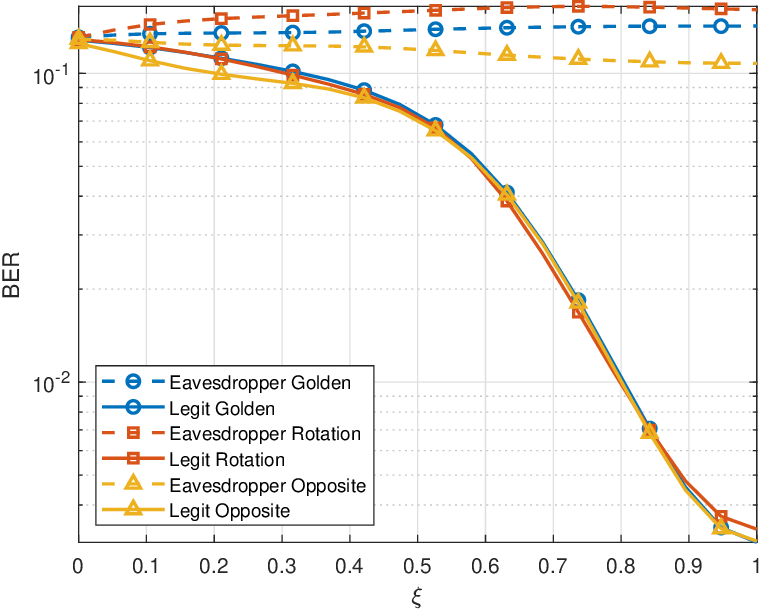}\label{fig:xpol_all_unbalance_16_15}}\hfill
	\caption{SNR degradation due to unbalanced polarization as a function of $\xi$ for constellations $M=8$ and $M=16$.}
	\label{fig:xpol_all_unbalance}
\end{figure}

Similarly, Fig. \ref{fig:xpol_all_unbalance} presents the degradation of SNR in the case of polarization imbalance. In contrast to cross-polarization, when $\xi\rightarrow 0$, it indicates maximum imbalance, where one polarization is blocked, leading to the highest level of degradation.

\section{CONCLUSIONS}
In conclusion, this paper provides a thorough exploration of secure systems operating at the physical layer, specifically utilizing the polarization properties of electromagnetic fields. The presented research establishes the necessary conditions for the feasibility of such systems and introduces innovative approaches for encryption and decryption processes directly on electromagnetic signals. By introducing novel metrics to measure strength and security, the paper enables the optimization of these systems, facilitating the identification of optimal solutions.

The study presents three distinct secure systems with different secret pattern designs, tailored to specific application scenarios. The investigation of imperfect polarization sheds light on its impact and provides valuable insights into system performance in real-world conditions. The experimental results validate the effectiveness and robustness of the proposed systems, highlighting their practical viability for secure communication.

Overall, this work advances the field of physical layer security, offering a comprehensive framework for the design and implementation of secure communication protocols. The introduced metrics, optimized systems, and analysis of imperfect polarization contribute to a deeper understanding of the subject, paving the way for enhanced security in modern communication networks.


\printbibliography

\end{document}